\documentclass{article}
\pdfoutput=1 
\usepackage{geometry}
\usepackage{graphicx}
\usepackage{amsmath}
\usepackage{amssymb}
\usepackage[authoryear]{natbib}
\usepackage[mathlines]{lineno}

\newcommand{\Ra}{\text{Ra}}
\newcommand{\Nu}{\text{Nu}}
\newcommand{\chiglobal}{\langle\chi\rangle}
\newcommand{\bnabla}{\boldsymbol{\nabla}}

\begin{document}
%
%
%

\title{Mixing across fluid interfaces compressed by convective
  flow in porous media}

\author { Juan J. Hidalgo$^{1, 2}$ \& Marco Dentz$^{1,2}$\\
 \small{1 IDAEA-CSIC, Barcelona, 08034, Spain}\\
 \small{2 Associated Unit: Hydrogeology Group (UPC-CSIC), Barcelona, Spain} }

\maketitle
%
%
\begin{abstract}
  We study the mixing in the presence of convective flow in a porous
  medium. Convection is characterized by the formation of vortices and
  stagnation points, where the fluid interface is stretched and
  compressed enhancing mixing. We analyze the behavior of the mixing
  dynamics in different scenarios using an interface deformation
  model. We show that the scalar dissipation rate, which is related to
  the dissolution fluxes, is controlled by interfacial processes,
  specifically the equilibrium between interface compression and
  diffusion, which depends on the flow field configuration. We
  consider different scenarios of increasing complexity. First, we
  analyze a double-gyre synthetic velocity field. Second, a
  Rayleigh-B{\'e}nard instability (the Horton-Rogers-Lapwood problem),
  in which stagnation points are located at a fixed interface. This
  system experiences a transition from a diffusion controlled mixing
  to a chaotic convection as the Rayleigh number increases. Finally, a
  Rayleigh-Taylor instability with a moving interface, in which mixing
  undergoes three different regimes: diffusive, convection dominated,
  and convection shutdown. The interface compression model correctly
  predicts the behavior of the systems. It shows how the dependency of
  the compression rate on diffusion explains the change in the scaling
  behavior of the scalar dissipation rate. The model indicates that
  the interaction between stagnation points and the correlation
  structure of the velocity field is also responsible for the
  transition between regimes. We also show the difference in behavior
  between the dissolution fluxes and the mixing state of the
  systems. We observe that while the dissolution flux decreases with
  the Rayleigh number, the system becomes more homogeneous. That is,
  mixing is enhanced by reducing diffusion. This observation is
  explained by the effect of the instability patterns.
\end{abstract}
%
%
\section{Introduction}
Convective flow caused by an unstable stratification of fluid density
such as the Rayleigh-B{\'e}nard or the Rayleigh-Taylor instabilities
are common in porous media. The Rayleigh-B{\'e}nard instability
appears when an unstable density stratification is maintained between
the top and bottom boundaries of the domain. This is often found when
the fluid temperature is altered as in geothermal groundwater
systems~\citep{Cheng1979, Sanford1998} and heat conduction in metallic
foams~\citep{Dyga2015, Hamadouche2016}, or during the mixing of
freshwater and seawater in coastal aquifers~\citep{Cooper1964,
  Abarca2007}. The Rayleigh-Taylor instability occurs when one fluid
is placed on top of a less dense one. A situation found in geological
CO$_{2}$ storage~\citep{EnnisKing2005, Szulczewski2013}, the
displacement of dense contaminant plumes~\citep{Kueper1991}, or the
convection of compositional melts~\citep{Martin1987, Tait1989,
  Wells2011}. The coupling between flow and transport results in an
enhancement of boundary and dissolution fluxes and fluid mixing. Since
mixing leads to the attenuation of concentration contrasts and
dilution~\citep{Kitanidis:1994,DentzReview,LeBorgneDentzVillermaux2015}
and drives chemical reactions~\citep{DeSimoni2005,DentzReview},
understanding how unstable flow and mixing interact is therefore
essential to predict the behavior of such systems.

The behavior of mixing and dissolution fluxes is usually expressed in
terms of dimensionless quantities. In Rayleigh-B{\'e}nard
instabilities, the fluxes are represented by the Nusselt number
$\Nu$~\cite[see][]{Otero2004} and depend on the strength of the
instability given by the Rayleigh number $\Ra$. The numerical
simulations of \cite{Otero2004} found $\Nu \propto \Ra^{0.9}$ for
$1300 \lesssim \Ra \lesssim 10000$; \cite{Hewitt2012} found exponents
close to 1 for $\Ra > 1000$. These observations are in agreement with
the boundary layer analysis of \cite{Howard1966} that assumes that the
buoyancy flux is independent of the height of the domain. Although the
low $\Ra$ regime is not discussed, inspection of Figure 3 in
\cite{Otero2004} and Figure 2 in \cite{Hewitt2012} shows that
$\Nu \propto \Ra^{1/2}$ for $100 \lesssim \Ra \lesssim 1000$, which
differs from the $\Nu \propto \Ra^{2/3}$ predicted by
\cite{Kimura1986}.

Rayleigh-Taylor instabilities also display different scaling depending
on the dominant mechanism~\citep{Slim2014}. The system evolves from
$\Nu \propto \Ra^{1/2}$ when diffusion dominates to $\Nu \propto \Ra$
after the onset of the instabilities~\citep{Hidalgo2012,
Hidalgo2015}. In bounded domains, as the instabilities attenuate, the
relation between $\Nu$ and $\Ra$ becomes time
dependent~\citep{Hewitt2013a, Hidalgo2015}.

We focus on the behavior of dissolution fluxes, and fluid mixing in
the presence of convective flow. Mixing in unstable systems occurs at
the fluid interfaces that be located at the domain boundaries or at
another and whose shape is determined by the instability patterns. The
patterns organize themselves into cells, columnar plumes, or fingers,
and evolve jointly with the velocity field, which forms vortices and
stagnation points. The fluid interface is stretched and compressed at
these locations, especially at stagnation points, affecting the
magnitude of the fluxes across it.

We study the hydrodynamic mechanism of convective mixing and
dissolution and quantify them in an interface compression model that
is able to reproduce the observed mixing scaling. The model relates
the structure of the velocity field to mixing and dissolution fluxes
across the fluid interface. First, we present the governing
equations of the flow and transport in porous media and define the
observables that describe the system, namely, the scalar dissipation
rate and concentration probability density function and discuss their
relation to dissolution fluxes and mixing state of the system. Then we
introduce the interface compression model for the mixing and
dissolution fluxes in the vicinity of a stagnation point. We consider
three scenarios with increasing complexity. A double-gyre synthetic
velocity which is used to validate the interface compression model, a
heat transport problem in which a Rayleigh-B{\'e}nard instability is
triggered by the boundary conditions, and a two-fluid system in which
the density stratification provokes a Rayleigh-Taylor instability. The
interface compression models shows how mixing is controlled by the
structure of the velocity field, whose properties determine the
transition between scalings.
%
%
\section{Flow and transport governing equations}
Under the assumptions of incompressible fluids and the Boussinesq
approximation, the dimensionless governing equations for
variable-density single-phase flow in a 2D homogeneous porous medium
are~\citep{Riaz2006,Hidalgo2013}:
\begin{linenomath}
  \begin{align}
    &\bnabla\cdot\mathbf{q}=0,\label{eq:Flow}\\
    &\mathbf{q}=-\bnabla
p-\rho(c)\hat{\mathbf{e}}_{g},\label{eq:Darcy}\\
    &\frac{\partial c}{\partial t} + \mathbf{q}\cdot \bnabla c -
\frac{1}{\Ra} \nabla^{2} c =0, \label{eq:Transp}
  \end{align}
\end{linenomath}

where $p$ is a scaled pressure referred to a hydrostatic datum,
$\mathbf{q}$ is the dimensionless Darcy velocity, and
$\hat{\mathbf{e}}_{g}$ is a unit vector in the direction of
gravity. The dimensionless density $\rho$ is in general a function of
concentration $c$. Choosing as time scale the advective characteristic
time $t_{a} = L_{c}/q_{c}\phi$, where $L_{c}$ is the system length
scale, $\phi$ the porosity, and $q_{c} = k \rho_{c}g/\mu$ the
characteristic buoyancy velocity given by the permeability $k$,
viscosity $\mu$, a representative density $\rho_{c}$, and gravity $g$,
the transport equation~(\ref{eq:Transp}) is controlled only by the
Rayleigh number
\begin{linenomath}
  \begin{align}
    \label{eq:Ra} \Ra=\frac{q_{c}L_{c}}{\phi D_\text{m}},
  \end{align}
\end{linenomath}

where $D_\text{m}$~is the diffusion coefficient. The different scales
must be chosen depending on the problem solved and will be explained
when necessary.
 
The system behavior is analyzed in terms of the global scalar
dissipation rate
\begin{linenomath}
  \begin{align}
    \label{eq:chiglobal-def} \chiglobal = \frac{1}{\Ra}
\int_{\Omega}{d}\Omega\, |\bnabla c|^{2},
  \end{align}
\end{linenomath}

where $\Omega$ denotes the domain. At the steady state, $\chiglobal$
is equal to the flux through the boundaries \citep{Hidalgo2012} and
since $\Nu$ is defined as the that flux divided by the diffusive flux
over the domain ($1/\Ra$ in the current setup), it can be seen that
$\chiglobal =\Nu/\Ra$.

In closed systems the change of concentration
variance~\citep{LeBorgne2010} is equal to $2\chiglobal$. As the system
mixes and concentration homogenizes $\chiglobal$ goes to
zero. However, in the presence of sinks or sources the concentration
variance is also related to the boundary or dissolution
fluxes~\citep{Hidalgo2012}. In that case a non-zero $\chiglobal$
proportional to the fluxes can be found in the steady state. In that
case the mixing state of the system is better represented by the
probability density function (pdf) of the concentration calculated by
sampling the concentration in all the domain as
  \begin{align}
    \label{eq:pdf-def}
  p(c) = \frac{1}{A} \int\limits_{\Omega}
 \delta[c - c(\mathbf  x)] \,\mathrm{d} \Omega,
  \end{align}
 where and $A$ is the domain's area.

 The shape of the $p(c)$ when the system is well-mixed
 depends on the boundary conditions. For example, for a well-mixed
 closed system,  $p(c)$ is given by a Dirac delta
 centered at the average initial concentration.  If Dirichlet boundary
 conditions maintain a concentration difference between the system's
 boundaries and diffusion is the only transport mechanisms the
 concentration profile is linear and the pdf flat. Segregated systems
 are characterized by broad concentration pdfs with multiple local
 maxima.

To obtain information about the the spatial structure we shall use the
two-dimensional autocorrelation function
\begin{linenomath}
  \begin{align}
    \label{eq:acf} \text{ACF}_{g}(x,y) =
\mathcal{F}^{-1}\left\{\left|\mathcal{F}\{g(x,y)\}
\right|^{2}\right\},
  \end{align}
\end{linenomath}

where $g(x,y)$ whose autocorrelation is computed and $\mathcal{F}$
stands for the two-dimensional Fourier transform. The shape of ACF
indicates the presence of periodic structures. The correlation length
$l$ is related to the width of the first maximum of the ACF and gives
information about the size of those structures.

%
%
\section{Interface compression}
After the onset of instabilities, the fluid interface evolves under
the combined effect of velocity and diffusion~\citep{Elder1968}. In
the locations where the velocity field experiences sharp changes, such
as the stagnation points where the flow velocity goes to zero over a
distance equal to the interface thickness, the interface is compressed
and stretched. Diffusion, however, has the opposite effect and wants
to increase the interface width. The thickness $s$ of the interfacial
boundary layer thus is the result of the competition between
hydrodynamic compression and diffusive expansion, which can be
quantified by~\citep{Villermaux2012, LeBorgne2013}
\begin{linenomath}
\begin{align}
  \label{eq:s-evolution} \frac{1}{s} \frac{d s}{d t} = - \gamma +
\frac{1}{\Ra}\frac{1}{s^{2}},
\end{align}
\end{linenomath}
with the dimensionless compression rate $\gamma$; the dimensionless
diffusion coefficient is $\Ra^{-1}$. The compression rate is given by
the symmetric part of the strain tensor~\citep{Ottino1989}
\begin{linenomath}
  \begin{equation}
    \label{eq:strain-tensor-symm} \mathbf{E} = \frac{1}{2}(\bnabla
\mathbf{q} + \bnabla \mathbf{q}^{T}) =\left[
      \begin{array}{cc} \gamma & 0 \\ 0 & -\gamma
      \end{array} \right]
  \end{equation}
\end{linenomath}
The steady state solution of~\eqref{eq:s-evolution} determines the
length scale
\begin{linenomath}
\begin{align}
  \label{eq:sb} s_{B} = \frac{1}{\sqrt{\gamma\Ra}},
\end{align}
\end{linenomath}
at which the effects of compression and diffusion equilibrate. This
length is known as the Batchelor scale~\citep{Batchelor1959,
  Villermaux2006}.

In general the scalar transport in the vicinity of a stagnation point
located at a fluid interface can be described by the
advection-diffusion equation~\citep{Ranz1979, Villermaux2012,
  LeBorgne2013, Hidalgo2015}
\begin{linenomath}
\begin{align}
  \label{eq:tpt-stagpoint} \frac{\partial c}{\partial t} = \gamma z
\frac{\partial c}{\partial z} + \frac{1}{\Ra} \frac{\partial^{2}
c}{\partial z^{2}} ,
\end{align}
\end{linenomath}
where horizontal gradients are disregarded because they are small
along the interface and the stagnation point is located at
$z=0$. Following~\cite{Hidalgo2015}, the steady state solution for $c$
along its characteristics gives
\begin{linenomath}
\begin{align}
  \label{eq:diff-profile} c = c_{b} + \frac{1 - c_{b}}{2}
\text{erfc}\left(\frac{z}{\sqrt{2s_{B}^2}} \right),
\end{align}
\end{linenomath}
where it is considered that the concentration far above the
interface ($z \to -\infty$) is 1 and far below the interface
($z \to \infty$) has a value $c_{b}$, which can be different from
zero

Using~\eqref{eq:diff-profile} in~\eqref{eq:chiglobal-def} we obtain
the expression for $\chiglobal$
\begin{linenomath}
\begin{align}
  \label{eq:chiglobal-model} \chiglobal =
\frac{\omega_{e}}{\sqrt{4\pi}}\frac{ (1 - c_{b})^2}{s_{B} \Ra},
\end{align}
\end{linenomath}
where $\omega_{e}$ denotes an effective interface length in the
horizontal direction. The form of $\omega_{e}$ depends on the
characteristics of the flow and will be discussed for each of the
considered scenarios.
%
%
\section{Mixing around a stagnation point: the double-gyre}
To illustrate the interface compression model we analyze the behavior
of fluxes and mixing using a double-gyre velocity
field~\citep{Shadden2005}. This a simplified model of convective flow,
because flow and transport are uncoupled, and flow around stagnation
points. Similar models have been used to characterize mixing in
oceanic circulation~\citep{Musgrave1985}.
%
%
\subsection{Double-gyre}
We consider a rectangular domain of length 2 and height 1 in which the
incompressible velocity $\mathbf{q} = (q_{x},q_{z})$ is given by
\begin{linenomath}
  \begin{align}\label{eq:gyre-vel} q_{x} =& \sin(n\pi x)\cos(\pi z)/n
\\ q_{z} =& -\cos(n \pi x)\sin(\pi z),
  \end{align}
\end{linenomath}

where $n$ is a positive integer equal to 1 for the
double-gyre. Concentration is prescribed on top and bottom boundaries
so that
\begin{linenomath}
  \begin{equation} c(x,z) = \left\{
      \begin{array}{c} 0 \text{ at } z=0\\ 1 \text{ at } z=1
      \end{array} \right.
  \end{equation}
\end{linenomath}

and the lateral boundaries are periodic. Density is constant and flow
and transport are not coupled. There is no characteristic buoyancy
velocity, so we take $q_{c}=\max(q_{z})=1$ and
$\Ra = q_{c} L_{c}/\phi D_{m}$, which is in fact a P{\'eclet} number
since the velocity field is not related to convective instabilities.
%
%
\subsection{Interface compression and scalar dissipation}
\begin{figure} \centering
  \includegraphics[width=0.9\textwidth]{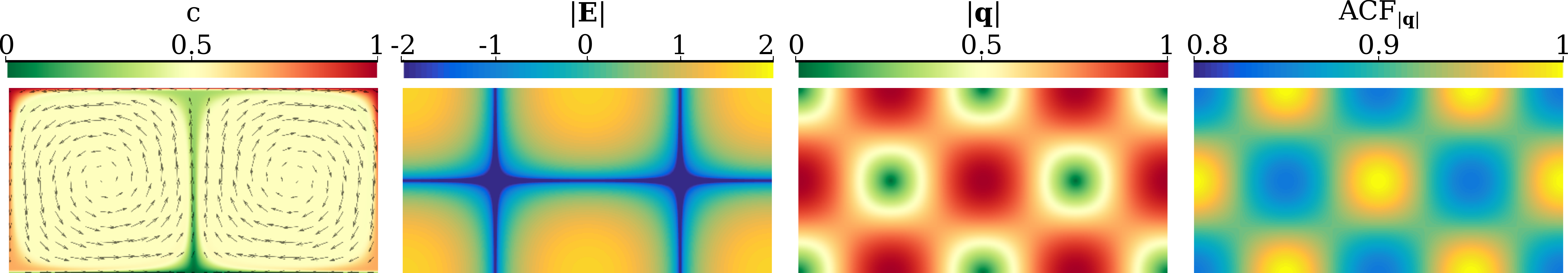}
  \caption{From left to right: steady state concentration and velocity
field (arrows) for the double-gyre $(\Ra=5000)$ , determinant of the
strain tensor $\mathbf{E}$, magnitude of the velocity, and 2D
normalized autocorrelation of the velocity field. The autocorrelation
is computed using the Wiener-Khinchin Theorem and results are shifted
so that the maximum is at the center of the domain.}
  \label{fig:gyre-strain}
\end{figure}

The velocity field varies smoothly along the vertical direction and
compresses the fluid against the top boundary and maintains the
concentration gradient. The structure of the velocity field can also
be visualized through the determinant of $|\mathbf{E}|$, which
displays extremes at the stagnation points.  The velocity 2D
autocorrelation also reveals the periodicity of the velocity
field~(Figure~\ref{fig:gyre-strain}).

There are eight stagnation points in the domain
(Figure~\ref{fig:gyre-strain}). At the steady state only the ones at
the boundaries contribute to mixing because the concentration
gradients inside the domain are zero. We take the one at the center of
the top boundary where the interface is compressed for the
calculations. The compression rate at that point is $\gamma = \pi$ and
\eqref{eq:sb} gives $s_{B} = \sqrt{1/\pi\Ra}$. Therefore from
\eqref{eq:chiglobal-model} we obtain
\begin{linenomath}
  \begin{align}
    \label{fig:chi-stag-point} \chiglobal \sim \frac{1}{2\sqrt{\Ra}}
  \end{align}
\end{linenomath}

The $\Ra^{-1/2}$ dependence was observed by~\cite{Ching2001} for
similar velocity fields.  In the double gyre, the velocity changes in
a scale of the order of the domain and $\gamma$ is therefore
independent of the value of $\Ra$. Thus the $\Ra^{-1/2}$ behavior is
characteristic of systems in which the velocity field (and $\gamma$)
and diffusion are uncoupled.

\begin{figure} \centering
\includegraphics[width=0.9\textwidth]{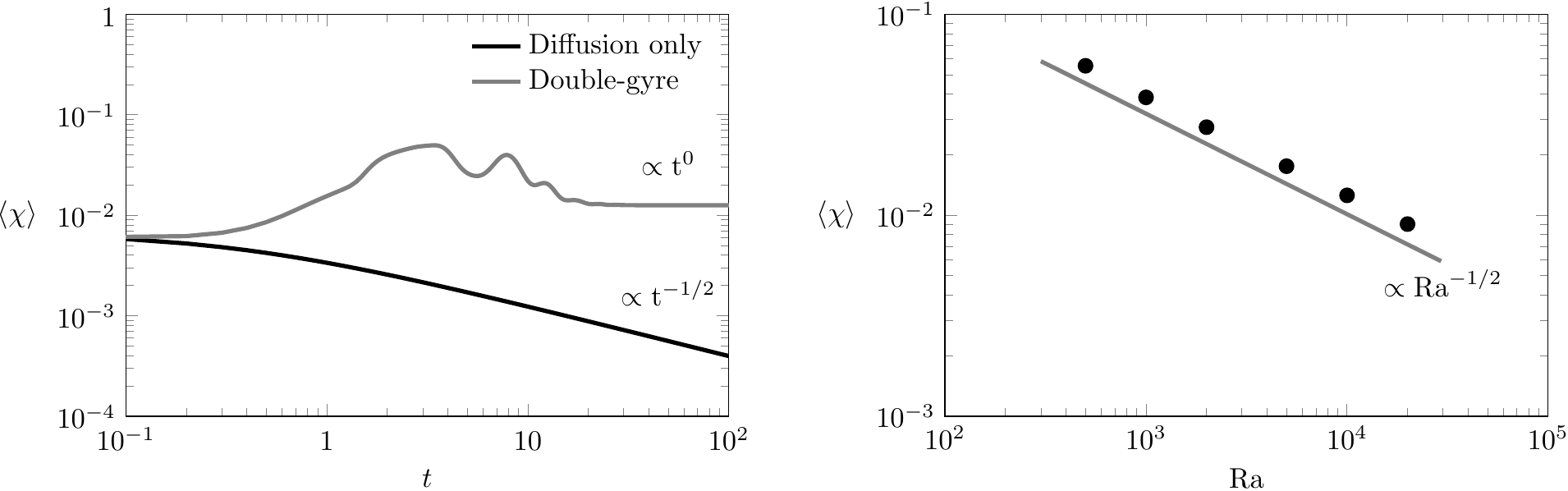}
\caption{Left: Comparison between a diffusion only case (black solid
  line), that is velocity equal zero, and the case with a double gyre
  velocity field for $\Ra=10000$ (gray solid line). The global scalar
  dissipation rate scales as $t^{-1/2}$ for late times in the
  diffusion case while the double-gyre evolves to a constant behavior.
  Right: Dependence of the total mixing $\chiglobal$ with $\Ra$ for
  the double gyre.}
  \label{fig:chi-gyre-Diff-comp}
\end{figure}

To verify the stagnation point model, we solved the double-gyre
transport problem for $500 < \Ra < 20000$. As expected, the effect of
the convection increases the mixing efficiency of the system
(Figure~\ref{fig:chi-gyre-Diff-comp} left), which arrives to a steady
state much faster than the only diffusion case. The global scalar
dissipation rate displays the expected $\Ra^{-1/2}$ behavior
(Figure~\ref{fig:chi-gyre-Diff-comp} right).

\begin{figure} \centering
  \includegraphics[width=0.3\textwidth]{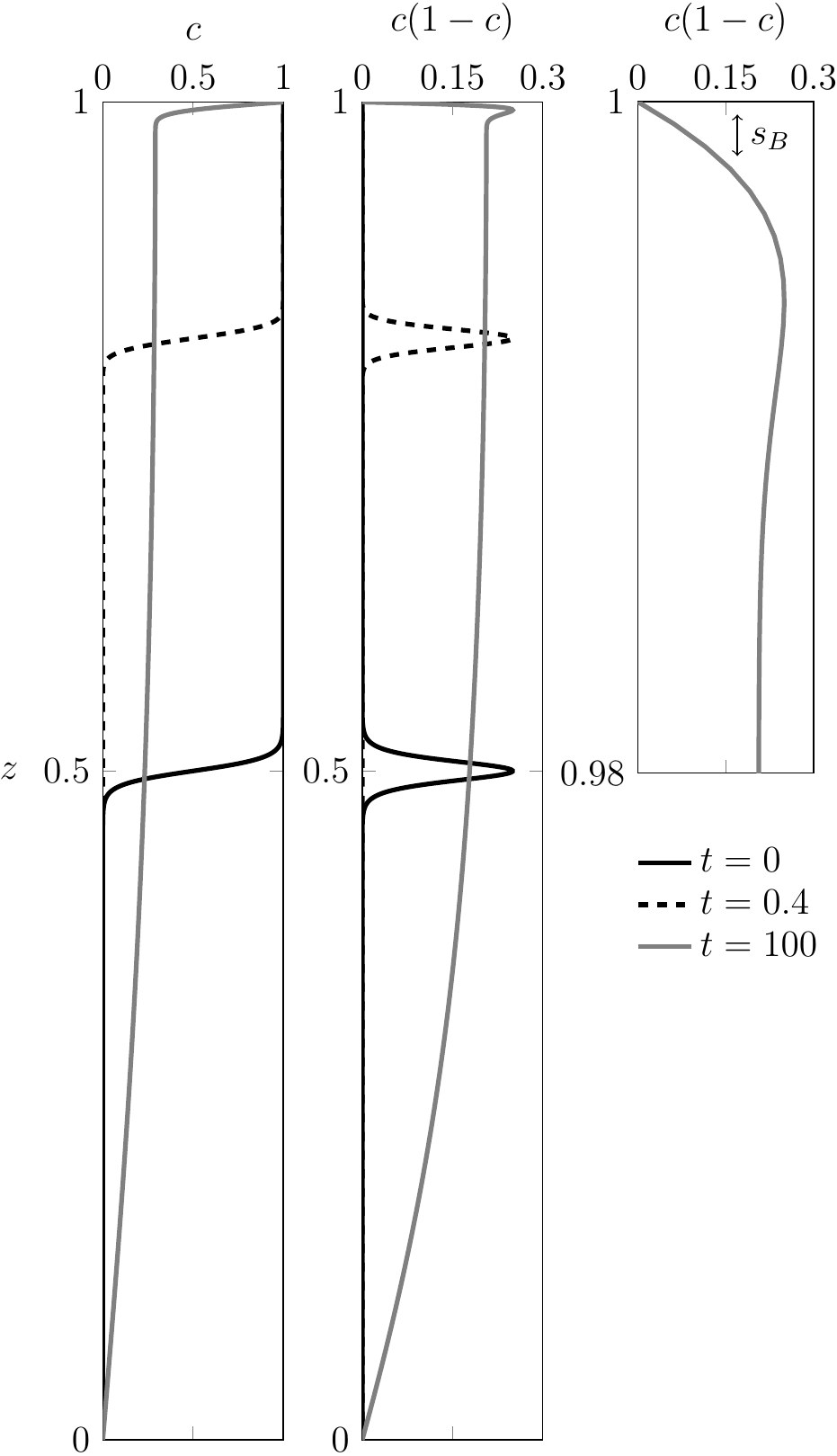}
  \caption{Concentration profile (left) and $c(1-c)$ of the
    concentration (right) for the double gyre ($n=1$) at $x=0.5$ for
    times $t= 0, 0.4, 100$. It can be seen how the interface is
    compressed against the top and bottom boundary as the system
    approaches to the steady state. The decrease in the interface
    width is shown by the shape of $c(1-c)$.}
  \label{fig:gyre-interface}
\end{figure}

\begin{figure} \centering
  \includegraphics[width=0.9\textwidth]{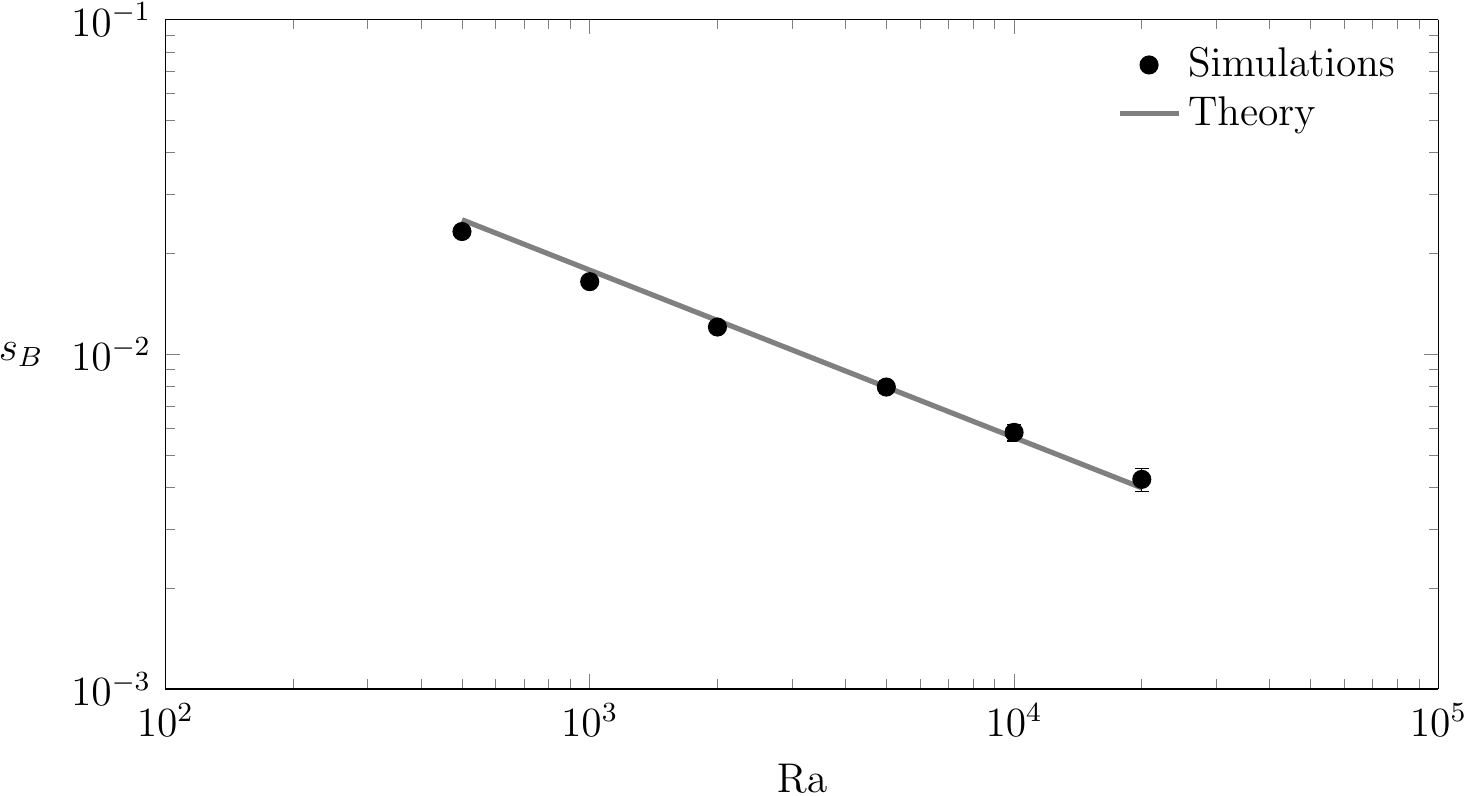}
  \caption{Computed (dots) interface width at the stagnation point and
theoretical (solid line) Batchelor scale ($s_{B} = (\pi\Ra)^{-1/2}$).}
  \label{fig:chi-gyre-analytic-num}
\end{figure}

As time passes the interface is compressed at the stagnation point
until the compression of the velocity field is balanced by diffusion
and the width equilibrates at the Batchelor scale $s_{B}$. The
interface width can be estimated as the square root of the second
central moment (variance) of $c(1-c)$ at the stagnation point as illustrated
in~Figure~\ref{fig:gyre-interface}.  There is a good agreement between
the theoretical Batchelor scale and the numerical
model~(Figure~\ref{fig:chi-gyre-analytic-num}).

\begin{figure} \centering
\includegraphics[width=0.9\textwidth]{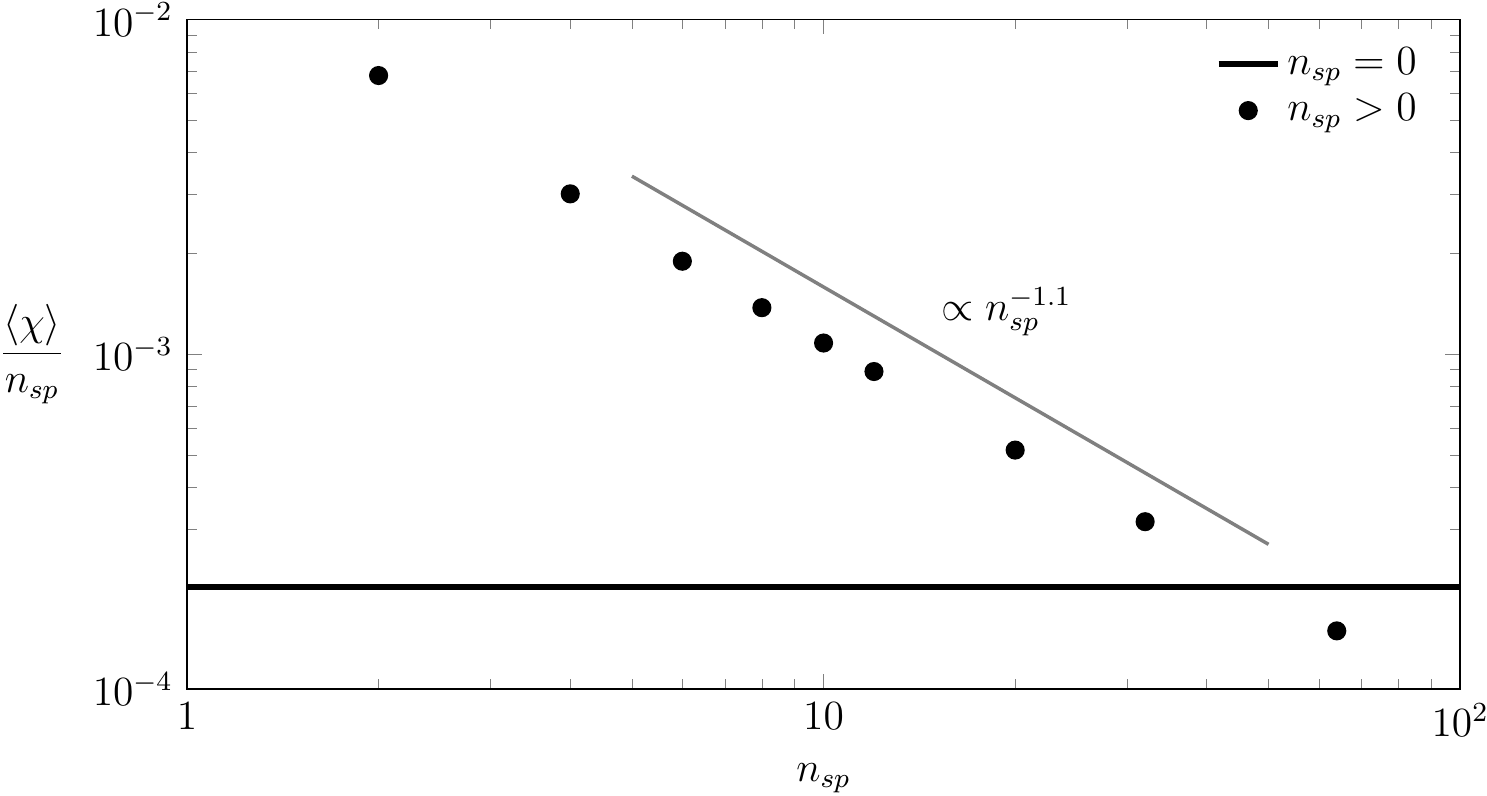}
\caption{Global scalar dissipation rate with increasing number of
  stagnation points for $\Ra =10000$. The solid horizontal line
  corresponds to the case in which diffusion is the only transport
  mechanism ($n_{sp} = 0$).}
  \label{fig:chi-ngyre}
\end{figure}

The number of stagnation points and convection cells in the system
increases with $n$. There are $2n-1$ convection cells and $2n$
stagnation points where the interface is compressed. The compression
rate at the stagnation points is independent of $n$ and so is
$\max(q_{z})$.  Therefore, $\Ra$ does not change. The simulations show
that $\chiglobal$ decreases with the number of cells
(Figure~\ref{fig:chi-ngyre} left). The decrease of dissolution
efficiency per stagnation point is caused by a reduction in the width
of the cells. This reduction behaves as $n^{-1}$ in good agreement
with the numerical results.
%
%
\subsection{Mixing state}
\begin{figure} \centering
\includegraphics[width=0.9\textwidth]{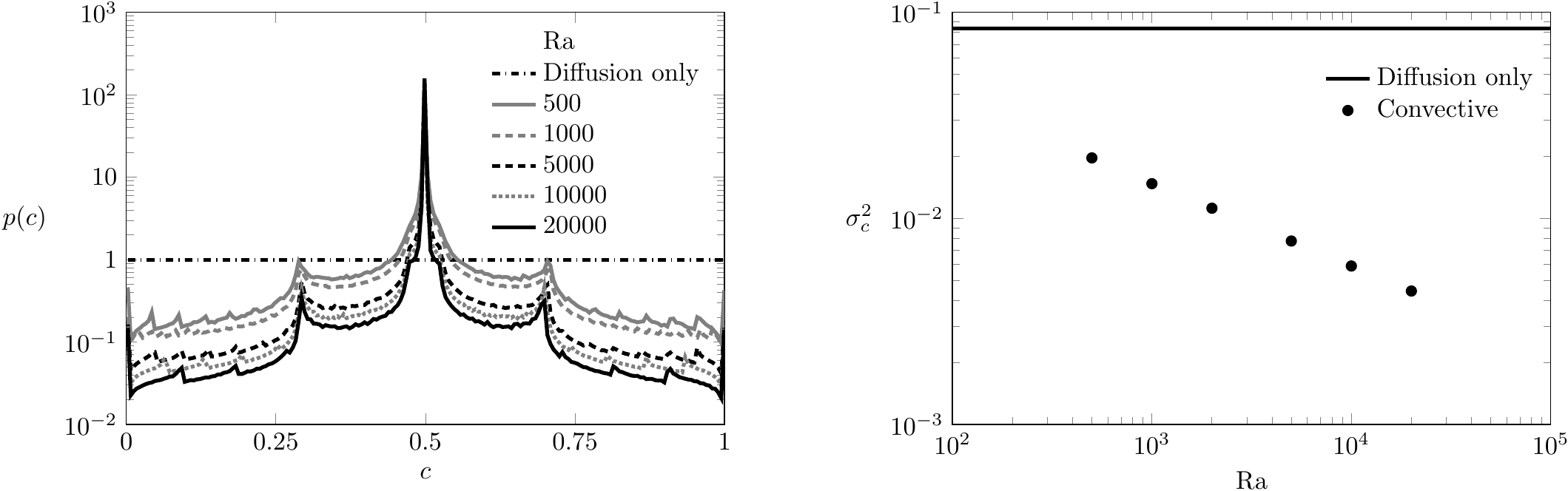}
\caption{Steady state concentration pdf for the double-gyre case
  ($n=1$) and different values of $\Ra$ (left) and the pdf variance
  (right). The diffusion only case is computed with $\Ra=10000$.}
  \label{fig:chi-hist}
\end{figure}

The mixing state of the system is represented by the concentration pdf
$p(c)$ and its variance $\sigma^{2}_{c}$ also show the effect of
convection. Without convection the pdf ~(Figure~\ref{fig:chi-hist}
left) is uniform for all the concentration range because the
concentration profile is linear. As $\Ra$ increases and the well-mixed
area inside the convection cells grows and then concentration
differences are confined near the boundaries. The weight of the
extreme concentration decreases and the pdf sharpens around the mean
value of $0.5$. The peaks at $c=0$ and $c=1$ corresponding to the
boundary conditions are always visible. The secondary peaks correspond
to the areas around the stagnation points near the boundaries where
the mean concentration is in between that of the boundary and the
well-mixed zone inside the cells.

Convection also helps in making the system more homogeneous. When
diffusion is the only mixing mechanism $p(c)$ has the maximum
variance possible because all values of concentration are equiprobable
(Figure~\ref{fig:chi-hist}~right). The concentration variance
$\sigma^{2}_{c}$ is inversely proportional to $\Ra$ reflecting the
above-mentioned reduction of the area occupied by concentration
gradients.

\begin{figure} \centering
\includegraphics[width=0.9\textwidth]{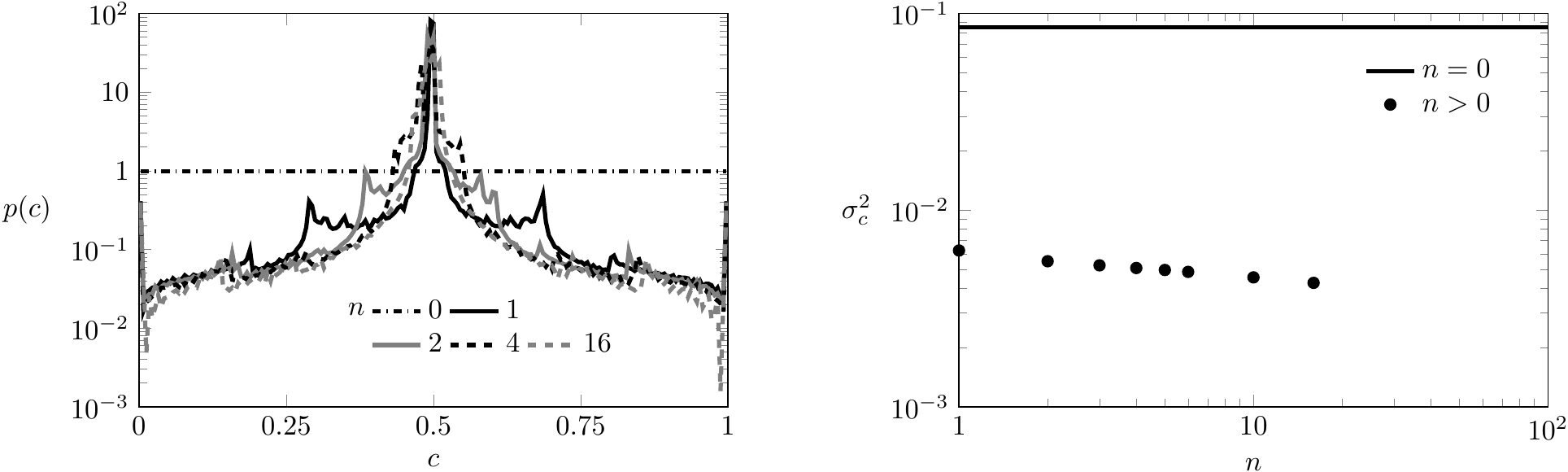}
\caption{Concentration pdf (left) and variance (right) for
  multiple-gyre case for different values of $n$ and $\Ra=10000$.  The
  velocity field homogenizes the system. However, the efficiency above
  $n=1$ decreases significantly.}
  \label{fig:ngyre-hist}
\end{figure}

The number of cells also affects the system state. The concentration
pdf displays a sharper shape (Figure~\ref{fig:ngyre-hist} left) and
the number of secondry peaks increases with $n$. The stirring of
additional convection cells, however, does not improve the homogeneity
of the steady state system significantly.  The variance of
concentration (Figure~\ref{fig:ngyre-hist} right) is not reduced
significantly by the addition of more cells. It is interesting to note
that the higher the dissolution flux, i.e. higher $\chiglobal$, the
less well-mixed the system is.

%
%
\section{Mixing across immobile interfaces}

We consider now a system for which the instabilities originate at the
boundary and propagate to the rest of the finite domain. The interface
is then fixed on one side and the shape of the instability patterns is
constrained in principle by the geometry of the system.
%
%
\subsection{The Horton-Rogers-Lapwood problem}

The Horton-Rogers-Lapwood (HRL) problem~\citep{Horton1945,Lapwood1948}
is a heat transport problem in which convection is triggered by a
Rayleigh-B{\'e}nard instability caused by the temperature difference
between the top and bottom boundaries. We solve the problem in a
rectangular domain of aspect ratio 2 (as in the double-gyre case) with
impervious top and bottom boundaries and periodic boundary conditions
on the sides. Temperature $T=1$ is prescribed at the top boundary and
$T=0$ at the bottom one. The dimensionless density of the fluid
increases linearly with temperature as $\rho = \beta T$, where $\beta$
is a positive constant. The system is again characterized by the
Rayleigh number~\eqref{eq:Ra}, which in this case takes the form $\Ra
= k \beta L_{c}/\phi\mu D_{m}$ since $q_{c} = k \beta / \mu$.

\begin{figure} \centering
\includegraphics[width=0.9\textwidth]{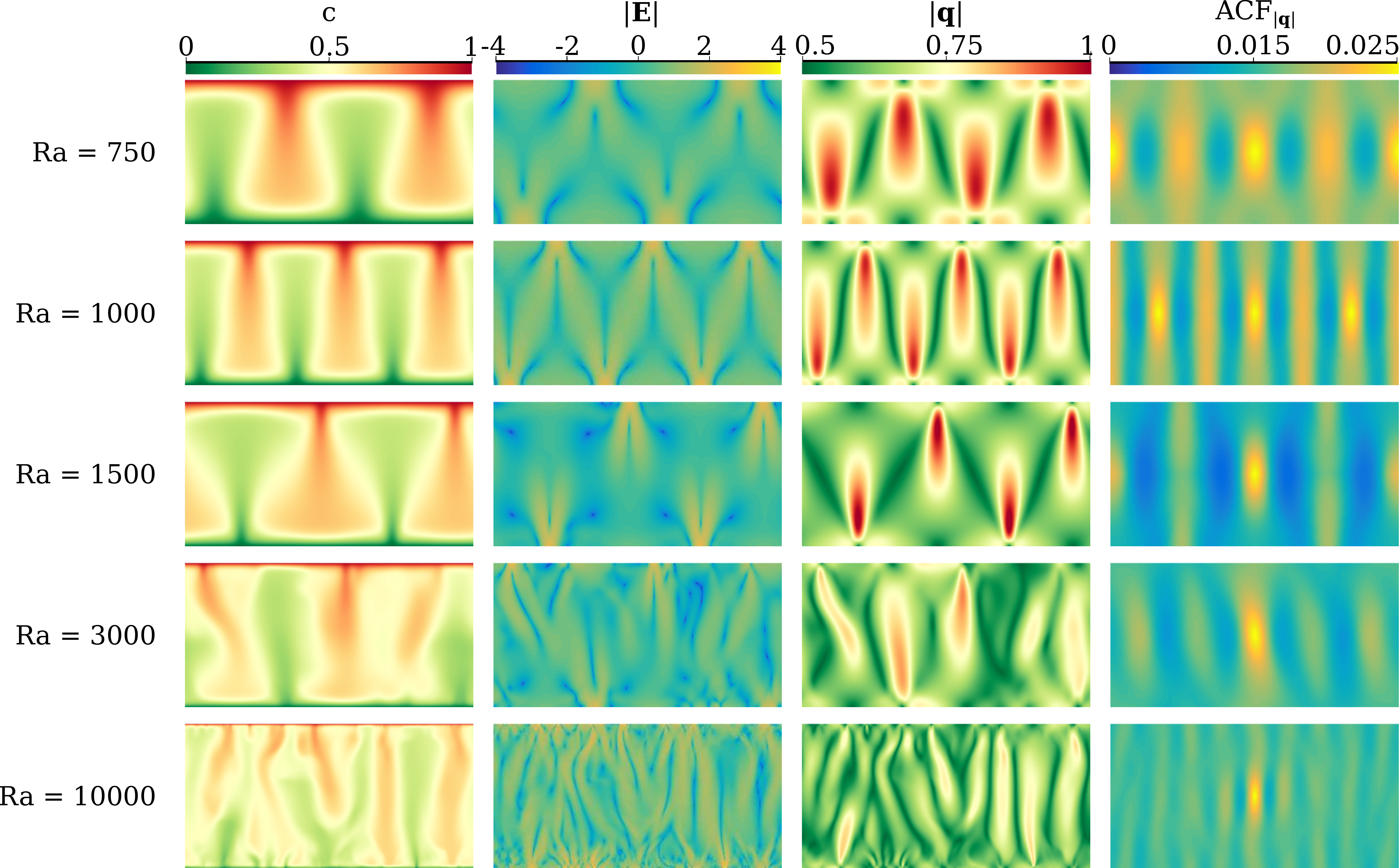}
\caption{From left to right: concentration, determinant of the strain
tensor $\mathbf{E}$, modulus of the velocity, and 2D normalized
autocorrelation of the velocity field for the Horton-Rogers-Lapwood
problem and different $\Ra$ at time $t=1000$. The autocorrelation is
computed using the Wiener-Khinchin Theorem and results are shifted so
that the maximum is at the center of the domain.}
  \label{fig:HRL-conc}
\end{figure}

The system is stable for $\Ra < 4\pi^{2}$. For $4\pi^{2} \lesssim \Ra
\lesssim 1300$~\citep{Graham1994} the instability patterns occupy the
whole domain in the form of convection cells
(Figure~\ref{fig:HRL-conc}). For higher $\Ra$ the system evolves to a
chaotic convection regime in which flow is organized in columnar
patterns~\citep{Hewitt2013b}. This transition occurs around $\Ra_{c}
\sim 1300$, which will be called critical Rayleigh number in the
following.

The stagnation points in this problem are located at the top and
bottom boundaries~(Figure~\ref{fig:HRL-conc}). For moderate $\Ra$ they
are found in between the convection cells and remain stationary once
the convection is fully developed. For high $\Ra$ when the system
experiences chaotic convection, the stagnation points are located at
the boundaries from where the columnar plumes grow. They appear and
disappear along the boundary as the small proto-plumes merge and
interact.
%
%
\subsection{Interface compression and and scalar dissipation}
\begin{figure} \centering
  \includegraphics[width=0.9\textwidth]{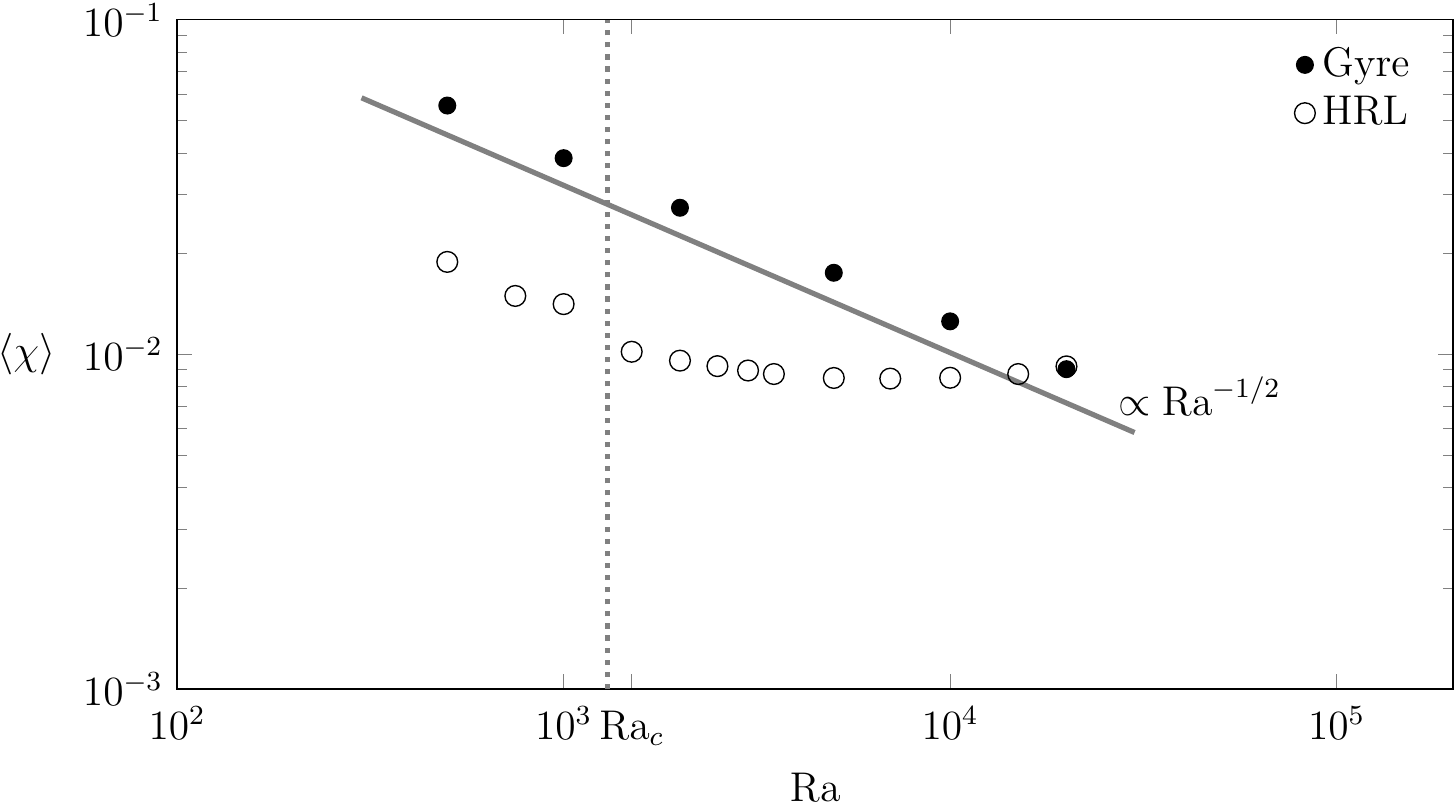}
  \caption{Dependence of the global scalar dissipation rate
with the Rayleigh number for the double gyre system and the
Horton-Rogers-Lapwood (HRL) problem. Mixing for HRL problem changes
its behavior from uncoupled ($\propto \Ra^{-1/2}$) to convective
around the critical Rayleigh number. This behavior has also been observed by \cite{Otero2004}  and \cite{Hewitt2012}.}
  \label{fig:chi-HRL}
\end{figure}

The global scalar dissipation rate $\chiglobal$ reflects the
transition of the system from an uncoupled, self-organized, convective
regime, in which $\chiglobal \propto \Ra^{-1/2}$ as in the double-gyre
scenario, to a convection dominated regime characterized by
$\chiglobal \propto \Ra^{0}$ (Figure~\ref{fig:chi-HRL}). The origin of
this change in the system's behavior lies in the structure of the
velocity field. As shown in Figure~\ref{fig:HRL-conc}, for low $\Ra$
the strain and the velocity field resemble that of the double-gyre as
the similarities in velocity and velocity autocorrelation
indicate. The velocity structure is dominated by the convection
pattern, which depends on the size and aspect ratio of the
domain. Therefore the velocity changes happen in the scale of the
domain size, as in the double-gyre case, and the compression rate is
independent of $\Ra$.  In the convection dominated regime, however,
the size of the domain becomes unimportant because the mixing process
happens at the scale of the interface, which is of the order of the
Batchelor scale. Velocity changes across a distance of the order of
$s_{B}$ and $\gamma$ grows linearly with $\Ra$. That is, compression
and diffusion become coupled.

\begin{figure}
  \centering
  \includegraphics[width=0.90\textwidth]{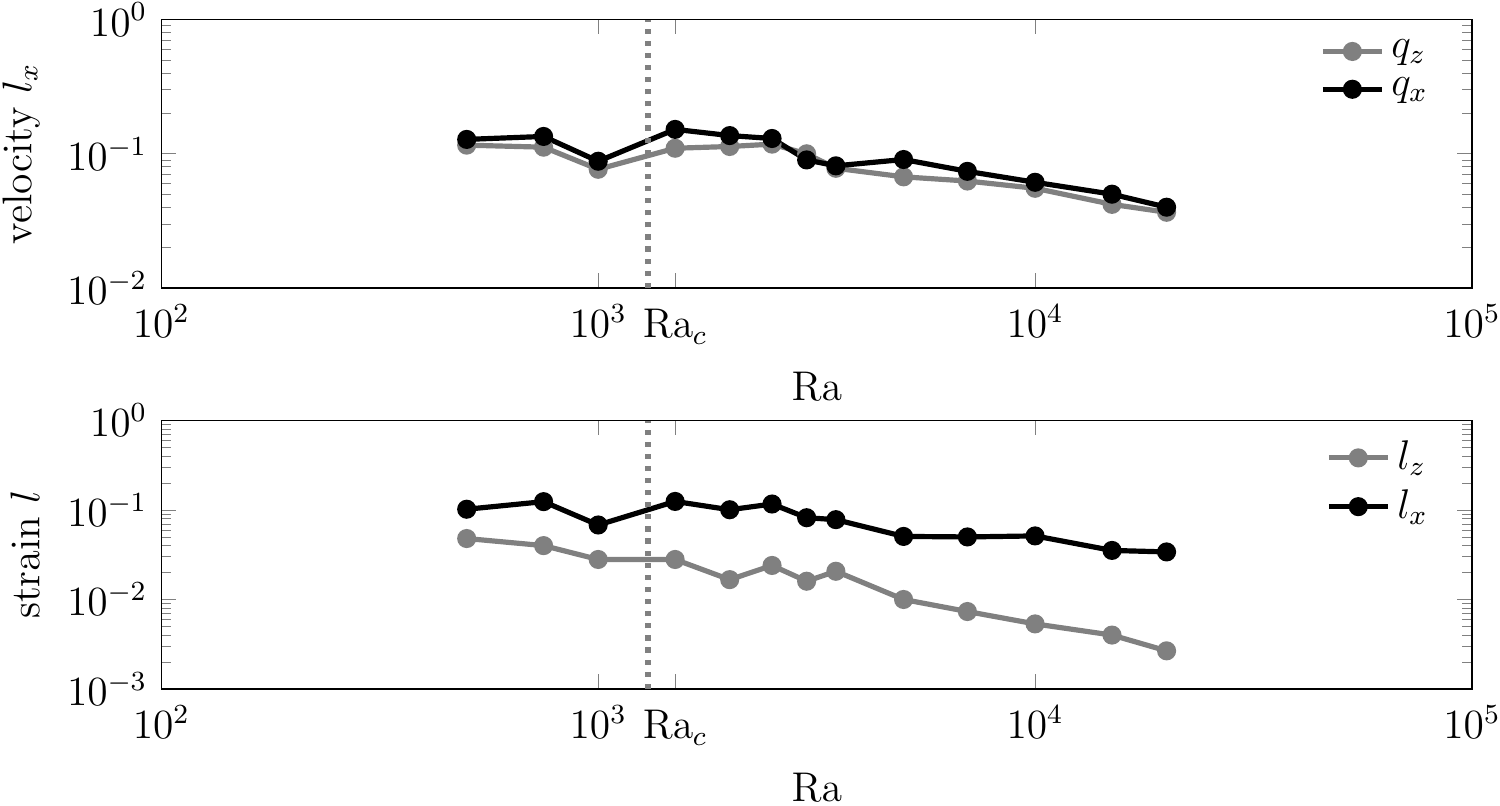}
\caption{Horton-Rogers-Lapwood problem velocity correlation length in
  the horizontal direction for both components of the velocity (top)
  and strain correlation length in both directions. The correlation
  length is computed using the averaged magnitudes from $t=20$ to
  $t=100$.}
  \label{fig:HRL-corr}
\end{figure}

This change in behavior is reflected in the correlation length in the
horizontal direction of the velocity and the strain. The correlation
length depends on the number of convection cells for $\Ra < \Ra_{c}$.
When a new cell is created as happens between $\Ra=750$ (2 cells, see
Figure~\ref{fig:HRL-conc}) and $\Ra=1000$ (3 cells) the correlation
length decreases (Figure~\ref{fig:HRL-corr}). It increases again when
$\Ra=1500$ and the system has two cells again. For $\Ra >\Ra_{c}$ the
correlation lengths decrease rapidly, indicating the transition to the
convective dominated regime.

The interface compression model
\eqref{eq:s-evolution}--\eqref{eq:chiglobal-model} explains the
observed behaviors of $\langle \chi \rangle$ in the different regimes
based on the scalings of the compression rate $\gamma$.  Regardless of
the regime, the difference in concentration across the interface at
the steady state is the one between the boundaries, that is
$c_{b} = 0$. The effective length $\omega_{e}$ associated to
the stagnation points during first regime is weakly dependent on $\Ra$
because it is linked to the number of convection cells, which
oscillates between 2 and 3 (Figure~\ref{fig:HRL-conc}). The
compression rate is, as explained before independent of
$\Ra$. Therefore from \eqref{eq:sb} and \eqref{eq:chiglobal-model} we
obtain $s_{B}, \chiglobal \propto \Ra^{-1/2}$.

During the convection dominated regime velocity changes sharply across
the interface thickness, $\gamma \sim 1/s_{B}$ and \eqref{eq:sb}
yields $s_{B} \sim \Ra^{-1}$. The effective length $\omega_{e}$ is
independent of $\Ra$ because it is proportional to the number of
stagnation points $(\sim \Ra)$ times their individual effective
length, which is proportional to the wavelength of the most unstable
mode $(\sim \Ra^{-1})$~\citep{Riaz2006,Hidalgo2015}. Therefore from
\eqref{eq:chiglobal-model} we obtain $\chiglobal \sim \Ra^{0}$.

\begin{figure} \centering
  \includegraphics[width=0.9\textwidth]{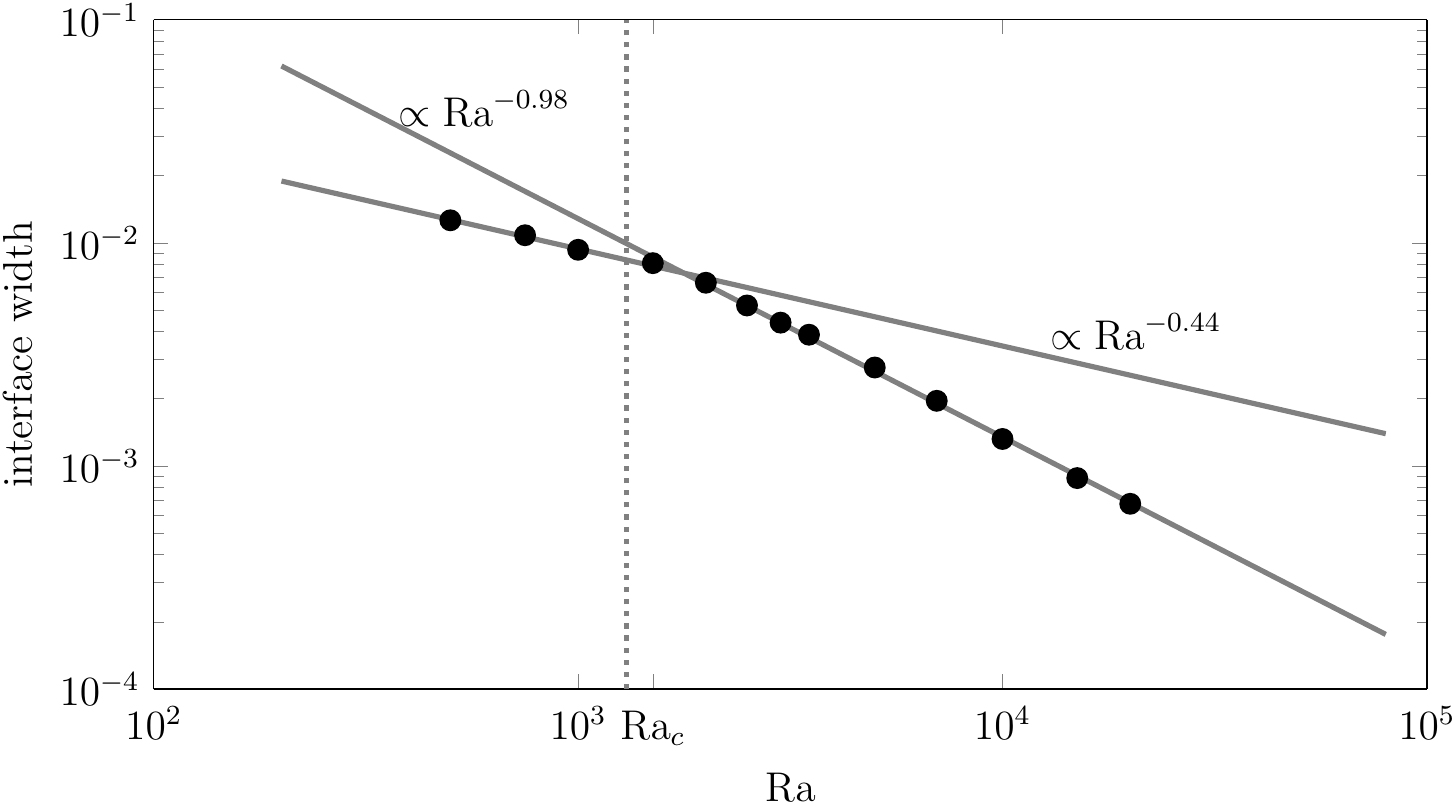} 
  \caption{Interface width dependence on $\Ra$ for the
    Horton-Rogers-Lapwood problem. The width is computed using the
    average concentration from $t=20$ to $t=100$ and it is as the
    square root of the minimum second central moment of $c(c-1)$ along
    the $x$ direction.}
  \label{fig:HRL-width}
\end{figure}

Numerical simulations confirm the former analysis. As in the
double-gyre case, we define the interface width as the square root of
the second central moment of $c(1-c)$. We compute the second central
moment at all locations along the top boundary and take as
representative the interface width the minimum measured value since
the movement along the boundary of the stagnation point and the
alternation of places where the interface is compressed leads to an
time average that overestimates the interface width. We observe
(Figure~\ref{fig:HRL-width}) that there is a change in the scaling of
the interface width around $\Ra_{c}$ from a value close to the
$\Ra^{-1/2}$ predicted by the model to a $\Ra^{-1}$ value virtually
equal to the model prediction and the observations in previous works~\cite{Rees2008, Hidalgo2009, Slim2010,Hewitt2013a}.
%
%
\subsection{Mixing state}
\begin{figure} \centering
  \includegraphics[width =0.9\textwidth]{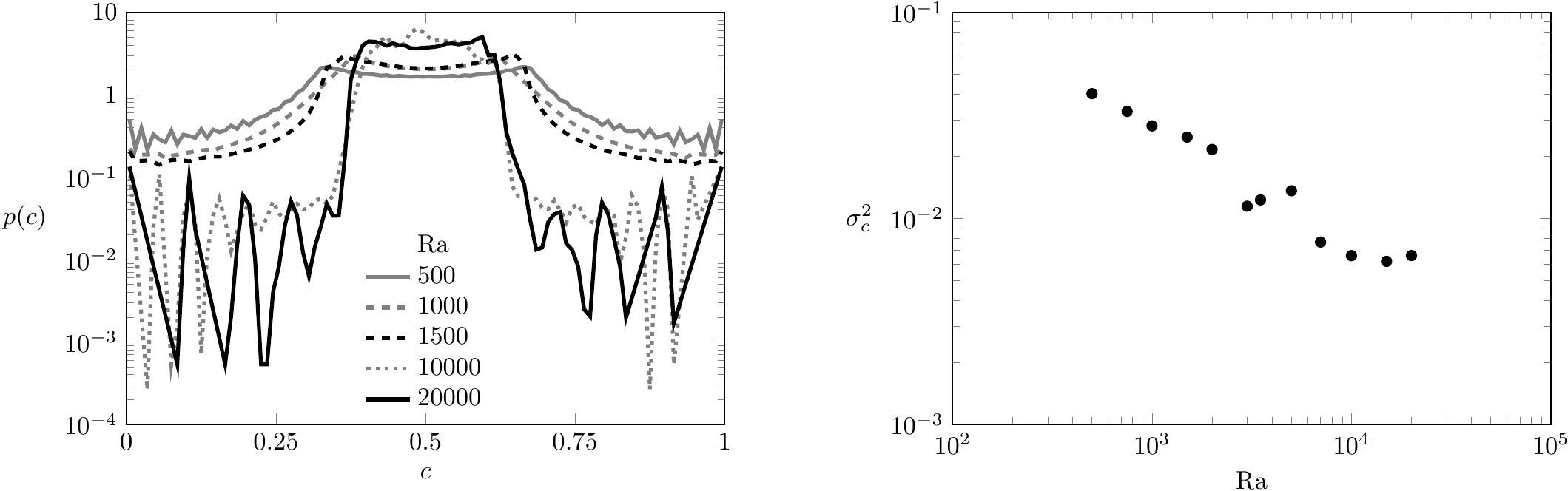}
  \caption{Probability density function of the time averaged
    concentration for the HRL problem (left) and pdf variance
    (right). Concentration was averaged from $t=20$ to $t=100$. Colors
    indicate different $\Ra$. For high $\Ra$ the pdf display
    some noise for the extrem values of concentration.}
  \label{fig:HRL-hist}
\end{figure}

Similarly to the double-gyre case the increasing strength of
convection narrows the concentration pdf around the average
concentration of $c=0.5$ (Figure~\ref{fig:HRL-hist} left). However,
$p(c)$ is not as sharp as in the double-gyre case because the area
between the convection cells or columnar plumes where the fluid is
well-mixed are smaller. This area grows as $\Ra$ increases, which
leads to a smaller concentration variance (Figure~\ref{fig:HRL-hist}
right) and a more homogeneous system.

Contrary to the scalar dissipation rate $\chiglobal$, which is
dominated by the concentration gradients at the interface, the mixing
state of the system does not become independent of $\Ra$ as the system
passes to the chaotic convection regime. The decrease in mixing
efficiency happens around $\Ra\approx 10000$, which is one order of
magnitude bigger than $\Ra_{c}$.

The dependence of $\sigma^{2}_{c}$ on $\Ra$ implies an in principle
counter-intuitive behavior: mixing increases with reducing
diffusion. The responsible for this behavior is the increasingly
chaotic convection, which stirs the system below the interface more
efficiently than the convection cells, and leads together with a
decreasing but finite diffusion to a more efficient homogenization of
the mixture.
%
%
\section{Stagnation points at mobile interfaces}
In stratified fluid systems the interface between the fluids is not
stationary and in general does not remain flat~\citep{Hewitt2013a}. We
relax now the assumption of a fixed flat interface and analyze a
system subject to a Rayleigh-Taylor instability triggered by an
unstable stratification of fluids. The interface compression and
mixing model for this system was previously developed by
\cite{Hidalgo2015}, which we further discuss below.

\begin{figure} \centering
\includegraphics[width=0.9\textwidth]{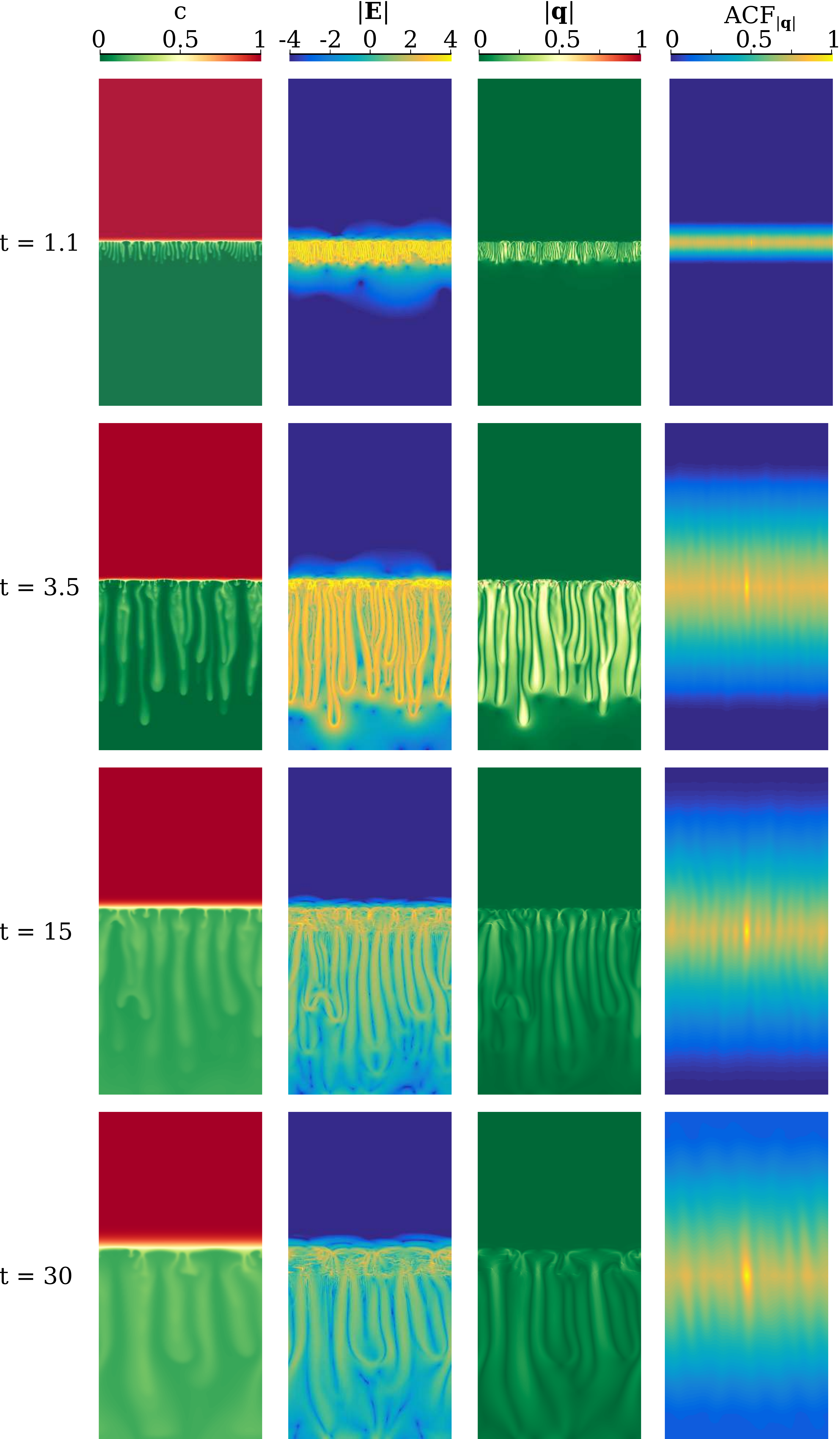}
\caption{From left to right: concentration, determinant of the strain
tensor $\mathbf{E}$, modulus of the velocity, and 2D normalized
autocorrelation of the velocity field for the two-fluid system under a
Rayleigh-Taylor instability for $\Ra=10000$ at different times.}
  \label{fig:convmix-strain}
\end{figure}
%
%
\subsection{Rayleigh-Taylor instability}
We consider a rectangular domain of length 1 and height 2 with top and
bottom impervious boundaries and periodic boundary conditions on the
sides~(Figure~\ref{fig:convmix-strain}). Initially the system is in
equilibrium with a less dense fluid on top of a dense one with the
interface located at $z=1$. The fluids are fully
miscible. Instabilities are triggered by a fluid non-linear
non-monotonic density law based on the mixtures of propylene-glycol
and water~\citep{Backhaus2011, DowChem} which is approximated in
dimensionless form by
$\rho(c) = 6.19c^{3} -17.86 c^{2} + 8.07c$~\citep{Hidalgo2015}. Note
that the dimensionless density is zero for $c=0$ (bottom fluid) and
negative for $c=1$ (top fluid). The maximum density is found at
$c_{m}=0.26$ so that the mixture of the fluids is denser than any of
the pure ones~\citep{Neufeld2010,Hidalgo2012,Hidalgo2015}. Again, the
system is completely characterized by the Rayleigh
number~\eqref{eq:Ra} defined with $q_{c} = k \Delta \rho g H_{0}/\mu$,
where $\Delta \rho$ is the density difference between the maximum and
the bottom fluid, and $L_{c} = H_{0}$ is the initial position of the
interface.
%
%
\subsection{Interface compression and and scalar dissipation}
\begin{figure} \centering
  \includegraphics[width=0.9\textwidth]{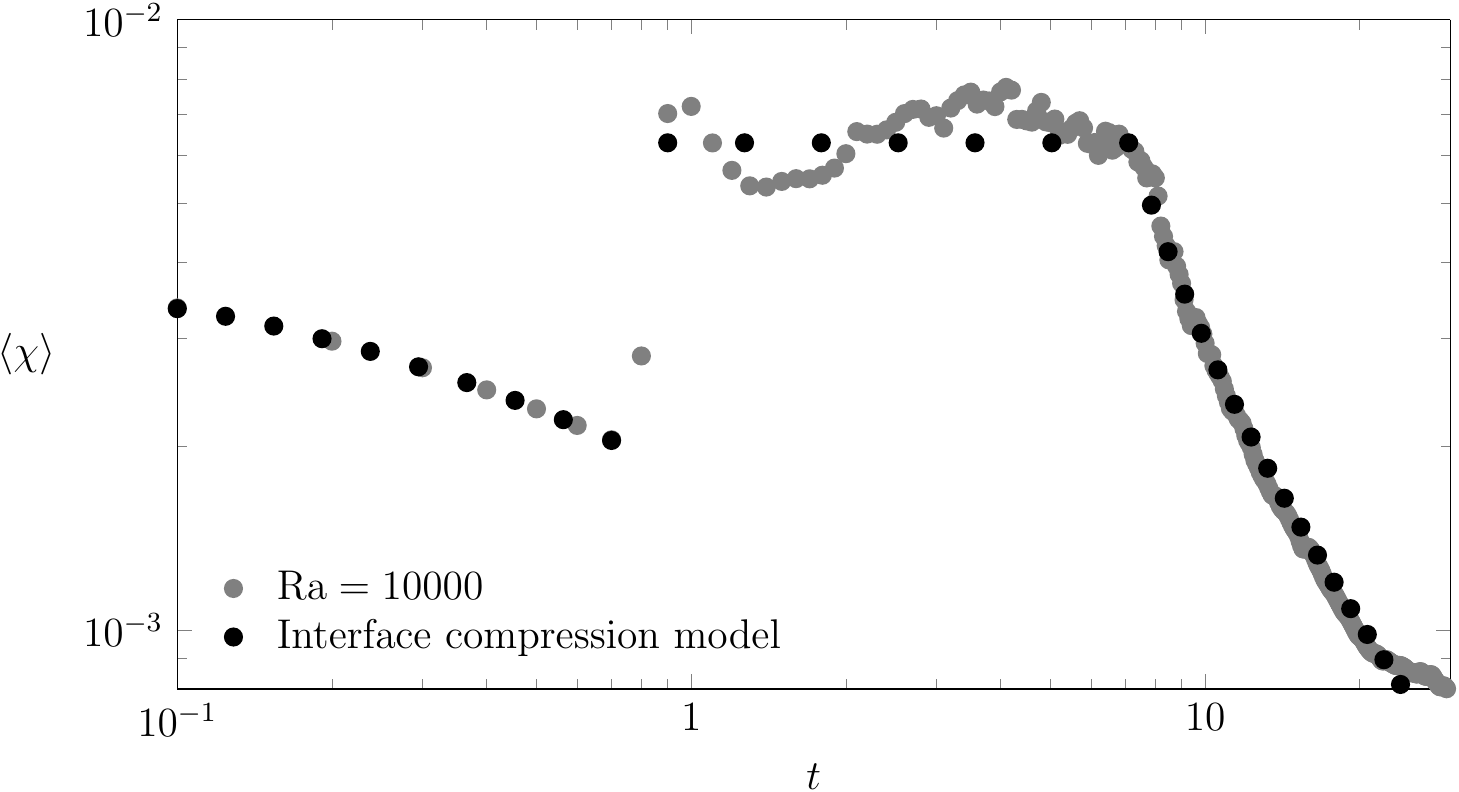}
  \caption{Evolution with time of the global scalar dissipation rate
for the two-fluid with $\Ra=10000$ for the numerical simulation (dots)
and the interface compression model (solid line).}
  \label{fig:convmix-chi}
\end{figure}

The global scalar dissipation rate (Figure~\ref{fig:convmix-chi})
shows that there are three main regimes: diffusive, convection
dominated, and convection shutdown. At the beginning the fluids mix
diffusively until the increase of density at the interface creates
instabilities that lead to a convection dominated regime. The
convection dominated regime is characterized by a constant global
scalar dissipation rate, and the formation of fingering patterns. As
the fluids mix and the concentration difference between the fluids
diminishes, convection and mixing slow down.

\begin{figure} \centering
    \includegraphics[width=0.9\textwidth]{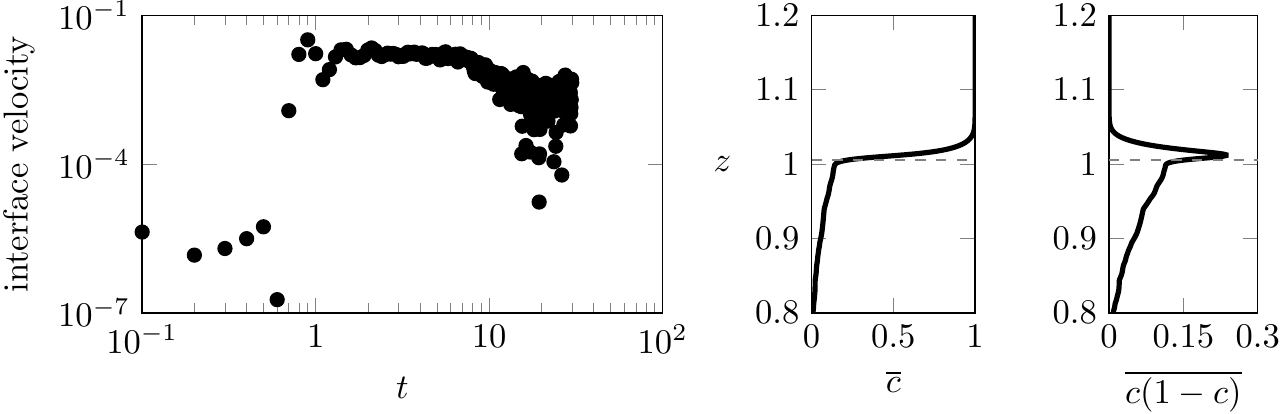}
    \caption{Interface velocity for the two-fluid system with $\Ra =
10000$ (left). Overlines indicate horizontally averaged
magnitudes. The interface position is determined as first zero of the
derivative of $\overline{c(1-c)}$ with respect to $z$ after the
maximum. This is illustrated on the plots on the right where the
$\overline{c}$ and $\overline{c(1-c)}$ are plotted for $t=1.4$,
$\Ra=10000$. The position of the interface is indicated by the black
dashed line.}
  \label{fig:ifc-vel}
\end{figure}

As in the previous problems, mixing is related to the interface and
velocity structure evolution. The main difference with the double-gyre
and the HRL problems is that the interface between the fluids is not
at rest as can be seen in the concentration maps in
Figure~\ref{fig:convmix-strain}. As the top fluid dissolves in the
bottom fluid the interface moves up. Figure~\ref{fig:ifc-vel} (left)
shows the velocity of the interface computed from $c(1-c)$ as
illustrated in the center and right panels of the same figure. The
maximum speed is observed during the convection dominated regime after
which the interface velocity decreases.

\begin{figure} \centering
  \includegraphics[width=0.9\textwidth]{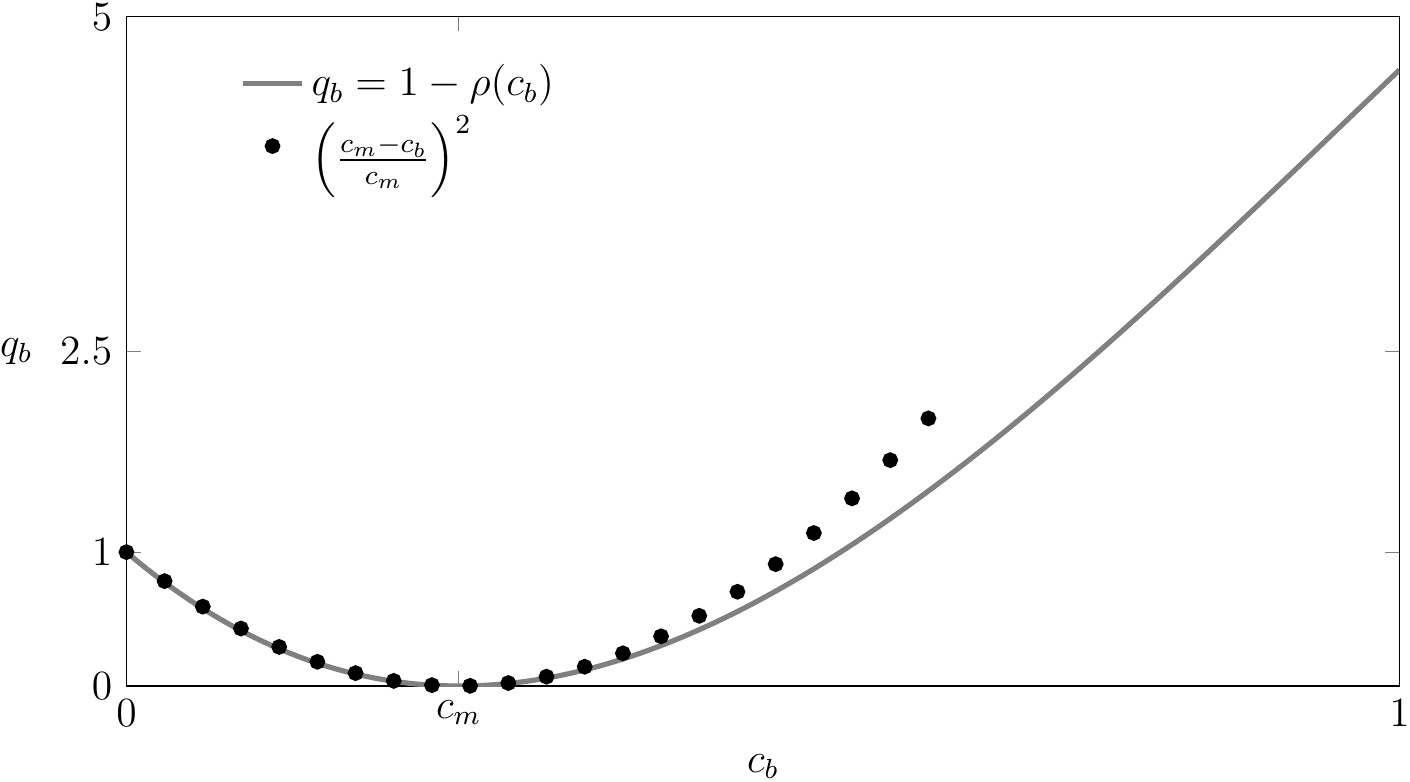}
  \caption{$\Delta\rho(c_{b}) = 1-\rho(c_{b})$ (solid line) can be
    approximated by $\left((c_{m} - c_{b})/c_{m}\right)^{2}$
    (dots). For the simulations in this study $c_{m}=0.26$.}
  \label{fig:parabola}
\end{figure}

The compression rate $\gamma$ is given by the net velocity change
across the interface as
\begin{linenomath}
  \begin{align} \gamma = \frac{q_{b} - q_{i}}{s},
  \end{align}
\end{linenomath}

where $q_{b}$ is the velocity of the up-welling fluid and $q_{i}$ is
the interface velocity at the stagnation point. The up-welling fluid
moves with a velocity proportional to the difference with respect to
the maximum density. For the chosen density law and $c_{b} < c_{m}$
$q_{b}$ can be approximated by~(Figure~\ref{fig:parabola})
\begin{linenomath}
  \begin{align} q_{b}(c_{b}) = \left(\frac{c_{m} -
c_{b}}{c_{m}}\right)^{2},
  \end{align}
\end{linenomath}

where, we recall, $c_{b}$ is the average concentration below the
interface and $c_{m}$ the concentration for which density is maximum.

The velocity of the interface at the stagnation point is proportional
to the dissolution flux, therefore
\begin{linenomath}
  \begin{align} q_{i} = \frac{1 - c_{b}}{s\Ra},
  \end{align}
\end{linenomath}

and the compression rate can be written as
\begin{linenomath}
  \begin{align}
    \label{eq:gamma-convmix} \gamma = \frac{1}{s}\left(q_{b} - \frac{1
- c_{b}}{s \Ra}\right).
  \end{align}
\end{linenomath}
Then the steady state solution of \eqref{eq:s-evolution} is
approximated by~\citep[see][]{Hidalgo2015}
\begin{linenomath}
  \begin{align}
    \label{eq:s-shutdown} s \approx \frac{2 - c_{b}}{(1 -
c_{b})^{2}\Ra}.
  \end{align}
\end{linenomath}

Equations \eqref{eq:gamma-convmix} and \eqref{eq:s-shutdown} show that
the fact that the interface motion leads to a lower compression of the
interface and smaller dissolution flux across it than in the HRL
problem. The difference in the dissolution flux between fixed and
moving interfaces was noted by \cite{Hidalgo2012}.

\begin{figure} \centering
\includegraphics[width=0.5\textwidth]{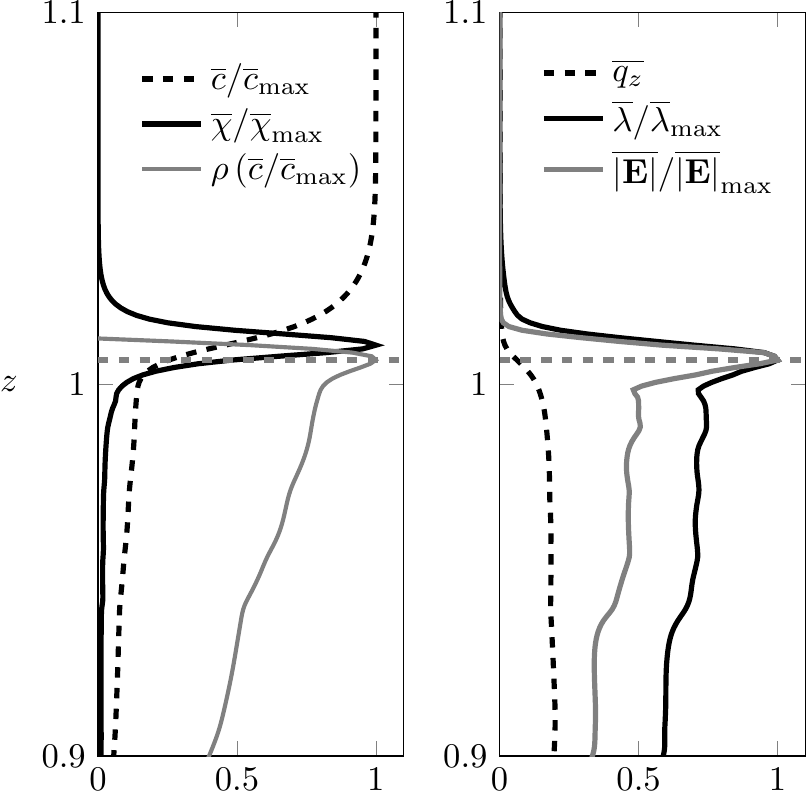}
\caption{Horizontally averaged concentration, scalar dissipation rate
  and density (left), and vertical component of the velocity,
  determinant of the strain tensor $\mathbf{E}$, and its maximum
  eigenvalue $\lambda$. Some magnitudes are normalized by their
  maximum value. Maximum mixing is found above the location of maximum
  compression (dashed line). Data correspond to the $\Ra=10000$ case
  for $t=1.4$.}
  \label{fig:c1c-strain}
\end{figure}

Maximum compression and scalar dissipation rate $\chi$ happen at the
interface where the maximum strain is also
found~(Figure~\ref{fig:convmix-strain}). There is also a high strain
on the sides of the fingers, however, their contribution to mixing is
small because the concentration difference between the finger center
and the surrounding fluid is low. Moreover, the increasing width of
the fingers softens the concentration gradient therefore reducing even
more their contribution to the mixing as time passes. The close
relation between velocity, mixing, and strain is illustrated in
Figure~\ref{fig:c1c-strain}. The maximum strain, indicated by the
determinant of the strain tensor and the maximum of its eigenvalues,
occurs at the same height as the maximum density. At this position the
derivative of $q_{z}$ is maximum too. Maximum scalar dissipation rate
however, is found a little above that point. This is caused by the
non-linear density law, which makes the interface asymmetric as it is
compressed only from the bottom. The interface asymmetry is more
severe in the HRL and double-gyre and problems because of static the
boundary (see Figure~\ref{fig:gyre-interface}).

When convection dominates, the up-welling fluid is still the initial
one, therefore $c_{b} = 0$ and $q_{b} = 1$. As a result the interface
compression is
maximum~(Figure~\ref{fig:width}). Using~\eqref{eq:s-shutdown} and
\eqref{eq:chiglobal-model} leads to
\begin{linenomath}
  \begin{align} s_{B} = \frac{2}{\Ra}
  \end{align}
\end{linenomath}
and
 
\begin{linenomath}
  \begin{align} \chiglobal =\frac{\omega_{e}}{4\sqrt{\pi}},
  \end{align}
\end{linenomath}

where $\omega_{e} \propto n_{sp}s_{B}$ with $n_{sp}$ the number of
stagnation points. At the onset of the instability, the fingers
distribute according to the wavelength $\lambda_{c}$ of the most
unstable mode. Therefore the $n_{sp}$ can be estimated using the
results of ~\cite{Riaz2006} as
$n_{sp} =1/\lambda_{c} = (\beta_{c} \Ra ) / (2\pi)$, which yields
$\omega_{e} = \beta_{c} ∕\pi$. Finally,
\begin{linenomath}
  \begin{align} \chiglobal = \frac{2}{\pi^{3/4}}\beta_{c},
  \end{align}
\end{linenomath}

which is independent of $\Ra$ because of the equilibrium between the
diffusive interface expansion and the compression exerted by the
buoyant fluid. \cite{Hidalgo2015} obtained $\beta_{c} = 0.018$ from
their simulations, which is $c_{m}$ times the one reported
by~\cite{Riaz2006} for a linear density law with $c_{m} = 1$.
Therefore the shape of the density law plays a critical role not only
in the location of the maximum
compression~(Figure~\ref{fig:c1c-strain}) but also on the value of the
scalar dissipation rate during convection.

\begin{figure} \centering
\includegraphics[width=0.9\textwidth]{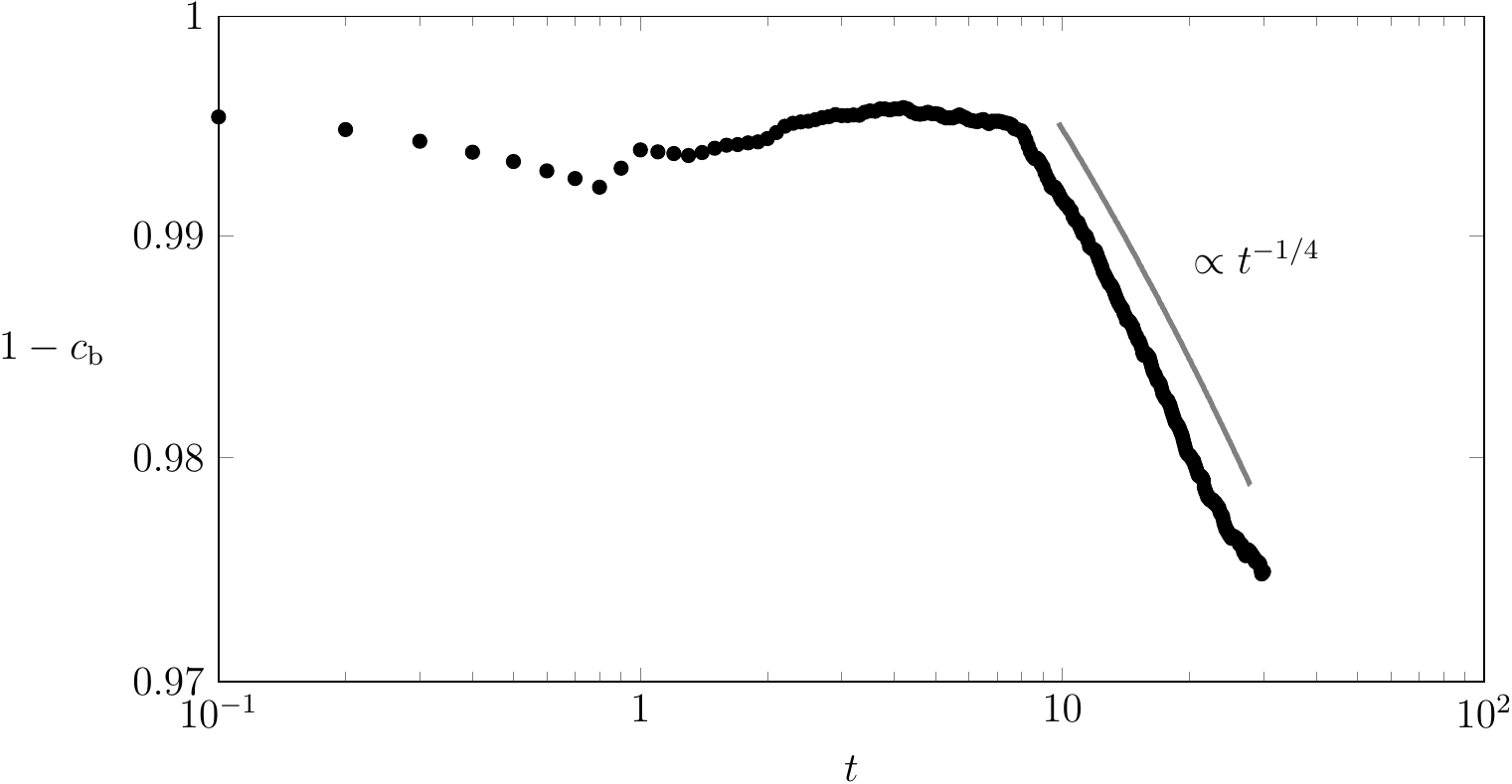}
\caption{Evolution of the average concentration below the interface
for the two-fluids system ($\Ra = 10000$). The concentration $c_{b}$
is constant during the convection dominated regime and decreases as
$t^{-1/4}$ during convection shutdown.}
  \label{fig:deltacb}
\end{figure}

\begin{figure} \centering
  \includegraphics[width=0.9\textwidth]{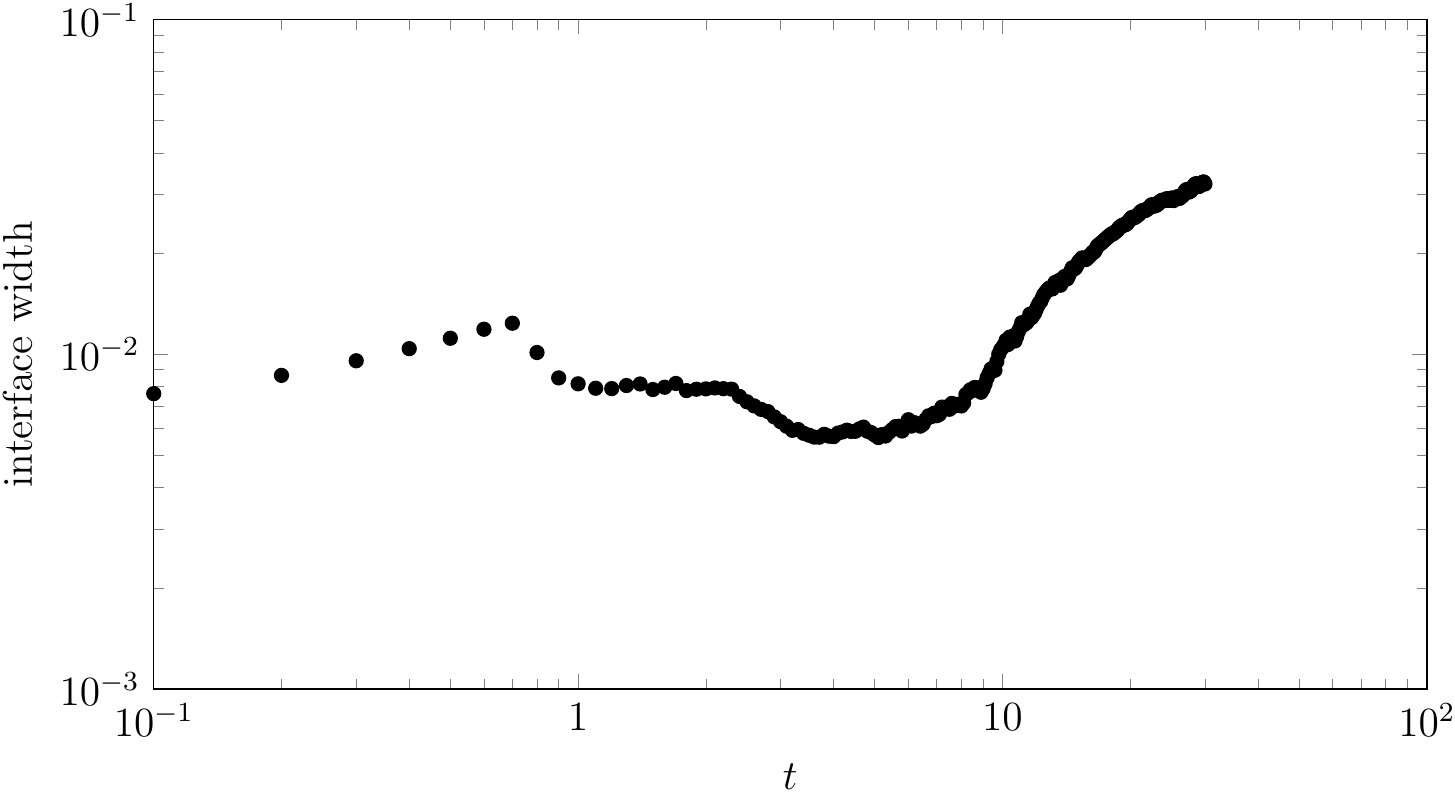}
  \caption{Interface width for the two fluids system ($\Ra = 10000$).
The interface width is defined as the square root of the second moment
of the horizontal average of $c(1-c)$. This value follows the expected
temporal evolution but overestimates the value of the interface width
because of the influence of the fingers (see
Figure~\ref{fig:ifc-vel}).}
  \label{fig:width}
\end{figure}

As the concentration of the bottom fluid increases the interface
compression and convection weaken and the interface width grows
rapidly as can be seen comparing figures~\ref{fig:deltacb} and
\ref{fig:width} in which the increase in $c_{b}$ happens at the same
time in which the interface width grows. The $1-c_{b} \sim t^{-1/4}$
behavior in Figure~\ref{fig:deltacb} is in good agreement with the
results of \cite{Hidalgo2015}
\begin{linenomath}
  \begin{align}
    \label{eq:cb-shutdown} c_{b} = 1 - \left[1 + 2 \omega_{e} (t -
\tau_{s})\right]^{-1/4}
  \end{align}
\end{linenomath}

which reproduces well the behavior of $\chiglobal$ during the
convection shutdown regime~(see Figure~\ref{fig:convmix-chi}). In
equation \eqref{eq:cb-shutdown} $\tau_{s}$ is the time when convection
shutdown begins and the effective length behaves as $\omega_{e}
\propto 0.002 \sqrt{\Ra}$ reflecting that the Rayleigh number becomes
meaningful again when the fingers reach the bottom of the system. From
that moment on velocity field is again influenced by the domain size
and the wide fingers behave similarly to the convection cells of the
HRL problem.

As the regimes succeed each other, the structure of the velocity field
changes~(Figure~\ref{fig:convmix-strain}). The maximum velocity is
found during the convection dominated regime and decreases as
convection shuts down. The velocity autocorrelation
ACF$_{|\mathbf{q}|}$ (Figure~\ref{fig:convmix-strain} right column)
reflects the horizontal structure of the fingering pattern with a
decreasing number of local maxima as fingers coarsen. During the
convection dominated regime the velocity and its autocorrelation are
similar to the high $\Ra$ HRL problem (compare $\Ra=10000$ in
Figure~\ref{fig:HRL-conc} to $t=3.5$ in
Figure~\ref{fig:convmix-strain}). During convection shutdown after the
fingers hit the bottom of the domain the velocity structure resembles
that of the low $\Ra$ HRL problem because the fingering patterns are
similar to elongated convection cells (compare $\Ra=750, 1000$ in
Figure~\ref{fig:HRL-conc} to $t=30$ in
Figure~\ref{fig:convmix-strain}).

\begin{figure} 
\centering 
\includegraphics[width =0.9\textwidth]{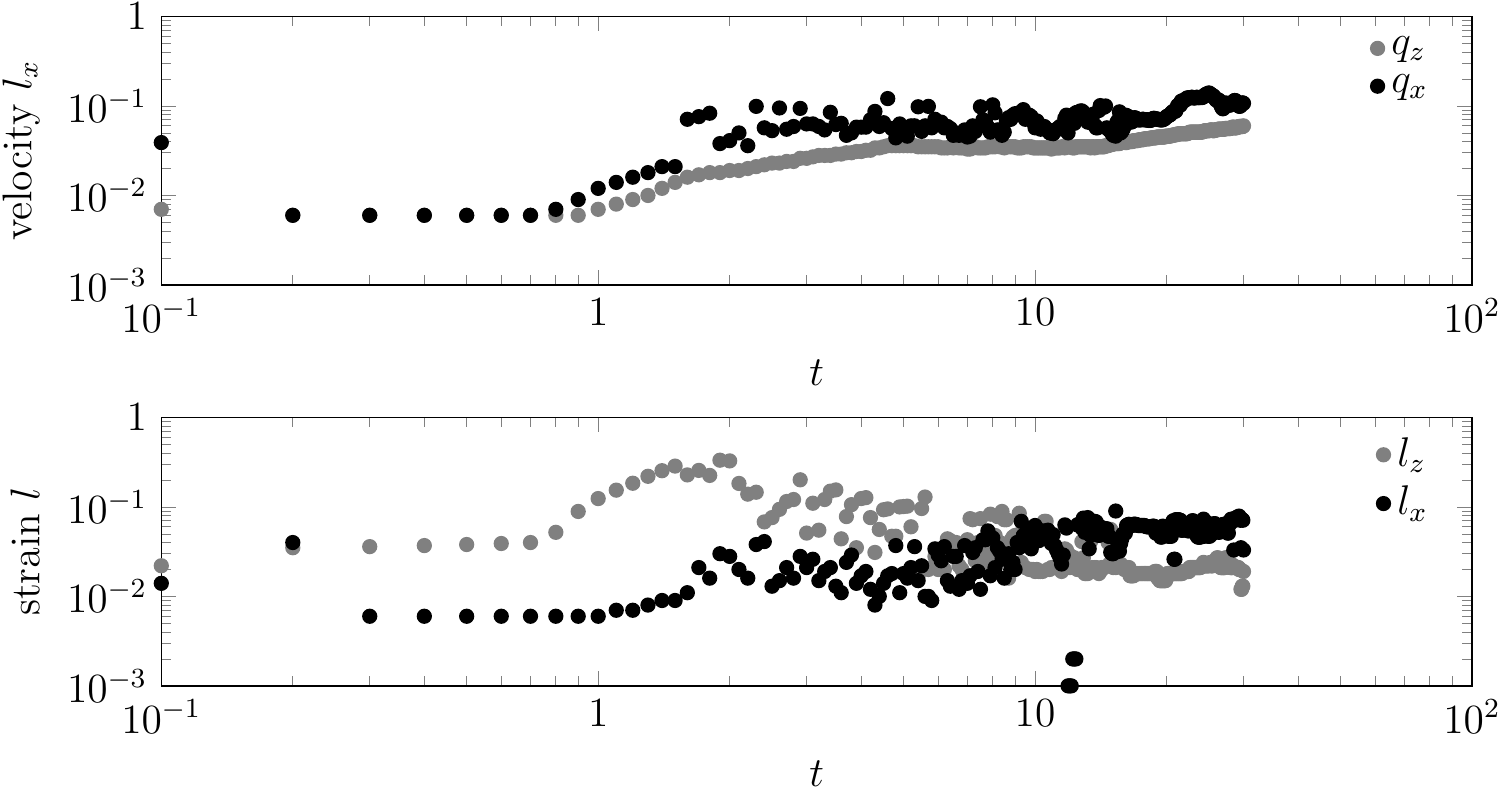}
\caption{Evolution with time of the velocity correlation length in the
horizontal direction for both components of the velocity (top) and
strain correlation length in both directions (bottom) for the
two-fluid system with $\Ra = 10000$.}
  \label{fig:convmix-corrlength}
\end{figure}

The velocity and strain correlation lengths also evolve during the
three regimes~(Figure~\ref{fig:convmix-corrlength}). The velocity
correlation length in both directions is minimum before the onset of
convection after which the maximum velocity is found. Then, the
velocity correlation length reflects the creation of the fingering
pattern. While the correlation length of $q_{z}$ grows and stabilizes
around a constant value, the correlation length of $q_{x}$ continues
growing as new fingers form and merge. A similar behavior is observed
for the correlation length of the strain. Its correlation length in
horizontal direction follows that of the horizontal velocity. In the
vertical direction, however, the maximum correlation length is found
after the onset of convection. This is caused by the growing fingers
along which there is a high strain. At late times it decreases as the
strain along the fingers becomes weaker.
%
%
\subsection{Mixing state}
\begin{figure} \centering
\includegraphics[width=0.9\textwidth]{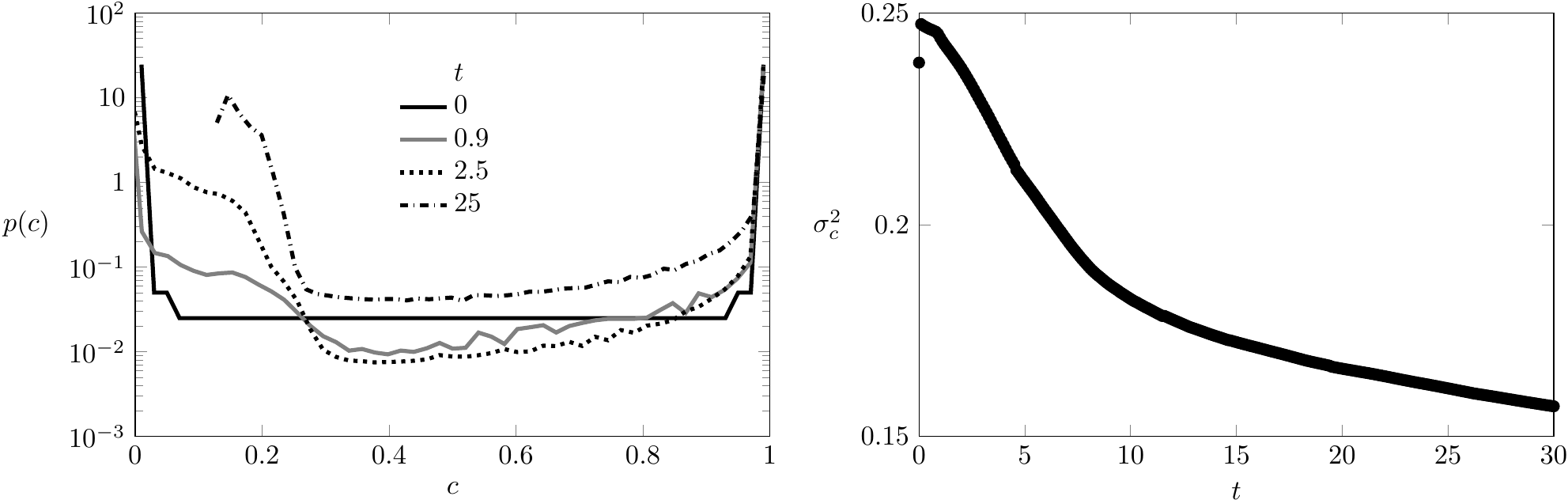}
  \caption{Concentration pdf for different times $\Ra=10000$.}
  \label{fig:convmix-hist}
\end{figure}

The mixing state of the system also changes with the different
regimes. In the beginning the system mixes slowly by diffusion and the
concentration pdf has two distinct peaks at the extreme concentrations
(Figure~\ref{fig:convmix-hist} left) has two distinct peaks at the
extreme concentrations. When convection takes over, the peak around
the low concentration shifts as a consequence of the mixing created by
the fingers. The peak around maximum concentration is widen by the
effect of diffusion. Eventually diffusion will take the system to a
well-mixed state with uniform 0.5 concentration because the fluid
occupied the same volume initially. However, there is an intermediate
state of duration of the order of $1/\Ra$ characterized by a skewed
concentration pdf displaying high probabilities around the
concentration of the initial top fluid ($c=1$) and a peak near $c_{m}$
as shown in Figure~\ref{fig:convmix-hist} (left) for $t=25$.

In the two-fluid system there are no boundary dissolution fluxes,
therefore, $\chiglobal$ is proportional to the time derivative of the
concentration variance. Figure~\ref{fig:convmix-hist} (right) shows
how the system homogenization evolves in accordance with
$\chiglobal$. Initially, the mixing state is given by the initial
conditions. During the onset of the instabilities $\sigma^{2}_{c}$
increases. However, it reduces as soon as they are fully
developed. This shows that the chaotic convection that creates the
fingering structures is an efficient mixing mechanism. As convection
shuts down, the bottom fluid mean concentration approaches $c_{m}$ for
which density is maximum and the density stratification approaches to
a stable configuration.  Then convection weakens and the fingers merge
and become wider, which makes the gradients of concentration at the
interface and below smaller. In this regime the mixing efficiency
decrease as well as the speed at which the system evolves towards the
well-mixed state.
%
%
\section{Conclusions}
We have studied the mixing in porous media under unstable flow
conditions using an interface compression model that is able to
reproduce the observed dependency of the global scalar dissipation
rate $\chiglobal$ (equivalent to the Nusselt number). The model,
introduced through the analysis of a problem with a synthetic
double-gyre velocity field, links the dissolution fluxes to the
interface width. The width of the interface is modified by the
velocity field, whose characteristics are related to the kind of
instabilities and the concentration evolution.

The Horton-Rogers-Lapwood problem was used to study a
Rayleigh-B{\'e}nard instability in which the fluid interface is
immobile. The system displays two regimes. First, it organizes itself
into convection cells as in the double-gyre. During this regime the
velocity field is not independent of the domain size and the
compression rate $\gamma$ is independent of diffusion, which leads to
$\chiglobal \propto \Ra^{-1/2}$. Above $\Ra_{c}$, the convection cells
turn into columnar plumes. The velocity autocorrelation decreases
abruptly and the system's size becomes irrelevant so that $\gamma$ is
of the order of $1/s_{B}$, therefore related to diffusion, and
$\chiglobal \propto \Ra^{0}$.

The case in which the interface is mobile was analyzed using a
Rayleigh-Taylor instability in which the unstable density
stratification was achieved by the mixture of two fluids with a
non-monotonic density law. The system experiences three regimes. A
diffusive regime in which the interface between the fluids grows. Then
a convection dominated regime after the onset of the instabilities in
which $\chiglobal$ is independent of $\Ra$. This behavior is similar
to the high $\Ra$ HRL problem. The domain size does not affect the
buoyancy fluxes and the interface width is controlled by a compression
rate linked to the diffusion. Finally, a convection shutdown regime in
which the system slowly approaches to a stable density stratification
as the fluids mix. This regime is characterized by a temporal
dependency of $\chiglobal$. The system then behaves as finite and the
correlation length of the velocity grows. 

The interface compression model and the analysis of the velocity field
revealed that the scaling of $\chiglobal$ is linked to the sytem size
experienced by the velocity field. When the velocity field and
concentration patterns are constrained by the domain boundaries,
$\chi \propto \Ra^{-1/2}$. However, when the structure of the velocity
field breaks because of the strong convection the size of the domain
becomes unimportant and $\chiglobal$ independent of $\Ra$.

We have shown that the global scalar dissipation $\chiglobal$ is
controlled by the dynamics of the fluid interface around the velocity
field stagnation points.  It is therefore expected that the stagnation
points play an central role in the location and magnitude of mixing
induced chemical reactions chemical reactions. The reaction hot spots
will be preferentially found near the maximum dissolution (and maximum
local scalar dissipation rate) takes place. The fingering and columnar
patterns contribute much less to $\chiglobal$. However they are
essential for the mixing state of the system.

The mixing state of the systems also depended on the nature of the
instabilities. The variance of concentration decreases by the mixing
of the convection patterns increases because of the fluxes through the
boundaries. The double-gyre and HRL problems reach a steady mixing
state in which both effects equilibrate and the variance of the
concentration remains constant. In both cases convection makes the
system more homogeneous. For low $\Ra$ the steady state is achieved
earlier and the dissolution fluxes are bigger because they are
proportional to $\Ra^{-1/2}$. For high $\Ra$ the system is better
mixed and displays a narrower concentration pdf but it takes it time
to arrive to that state. The Rayleigh-Taylor instability lacks
boundary fluxes. The evolution of the mixing state is governed by the
scalar dissipation rate. During the period in which convection
dominates mixing is maximum as well as the dissolution fluxes. As
convection ceases the efficiency of the system to mix itself
decreases. Therefore, the better mixed the system is the lower the
dissolution fluxes. This suggest that a certain level of segregation
might be desirable to maintain chemical reactions and fluxes through
the boundary. Contrary to intuition, the best mixing state, i.e.,
lower variance of concentration, is attained for high $\Ra$. That is a
reduction in diffusion favors the homogenization of the
concentration. This homogenization is achieved by the stirring created
by the instability patterns.

J.J.H. and M.D.  acknowledge the support of the European Research
Council through the project MHetScale
(FP7-IDEAS-ERC-617511). J.J.H. acknowledges the support of the Spanish
Ministry of Economy and Competitiveness through the project Mec-MAT
(CGL2016-80022-R).
%
%
\bibliography{bib}

\begin{thebibliography}{}

\bibitem[Abarca et~al., 2007]{Abarca2007}
Abarca, E., Carrera, J., S{\'{a}}nchez-Vila, X., and Voss, C.~I. (2007).
\newblock Quasi-horizontal circulation cells in 3d seawater intrusion.
\newblock {\em Journal of Hydrology}, 339(3-4):118--129.

\bibitem[Backhaus et~al., 2011]{Backhaus2011}
Backhaus, S., Turitsyn, K., and Ecke, R.~E. (2011).
\newblock Convective instability and mass transport of diffusion layers in a
  {H}ele-{S}haw geometry.
\newblock {\em Phys. Rev. Lett.}, 106(10).

\bibitem[Batchelor, 1959]{Batchelor1959}
Batchelor, G.~K. (1959).
\newblock Small-scale variation of convected quantities like temperature in
  turbulent fluid part 1. general discussion and the case of small
  conductivity.
\newblock {\em Journal of Fluid Mechanics}, 5(1):113--133.

\bibitem[Cheng, 1979]{Cheng1979}
Cheng, P. (1979).
\newblock Heat transfer in geothermal systems.
\newblock {\em Advances in Heat Transfer}, 14:1 -- 105.

\bibitem[Ching and Lo, 2001]{Ching2001}
Ching, E. S.~C. and Lo, K.~F. (2001).
\newblock Heat transport by fluid flows with prescribed velocity fields.
\newblock {\em Physical Review E}, 64(4).

\bibitem[Cooper, 1964]{Cooper1964}
Cooper, H.~H. (1964).
\newblock Sea water in coastal aquifers.
\newblock USGS Numbered Series 1613, USGS.

\bibitem[De~Simoni et~al., 2005]{DeSimoni2005}
De~Simoni, M., Carrera, J., S{\'a}nchez-Vila, X., and Guadagnini, A. (2005).
\newblock A procedure for the solution of multicomponent reactive transport
  problems.
\newblock {\em Water Resour. Res.}, 41(11):W11410.

\bibitem[Dentz et~al., 2011]{DentzReview}
Dentz, M., Le~Borgne, T., Englert, A., and Bijeljic, B. (2011).
\newblock Mixing, spreading and reaction in heterogeenous media: a brief
  review.
\newblock {\em J. Cont. Hydrol.}, 120-121:1--17.

\bibitem[{Dow Chemical}, 2011]{DowChem}
{Dow Chemical} (2011).
\newblock Propylene glycols - density values.
\newblock {https://dow-answer.custhelp.com/app/answers/detail/a\_id/7471}.

\bibitem[Dyga and Troniewski, 2015]{Dyga2015}
Dyga, R. and Troniewski, L. (2015).
\newblock Convective heat transfer for fluids passing through aluminum foams.
\newblock {\em Archives of Thermodynamics}, 36(1):139--156.

\bibitem[Elder, 1968]{Elder1968}
Elder, J.~W. (1968).
\newblock The unstable thermal interface.
\newblock {\em Journal of Fluid Mechanics}, 32(01):69 -- 96.

\bibitem[Ennis-King and Paterson, 2005]{EnnisKing2005}
Ennis-King, J.~P. and Paterson, L. (2005).
\newblock Role of convective mixing in the long-term storage of carbon dioxide
  in deep saline formations.
\newblock {\em {SPE} Journal}, 10(03):349--356.

\bibitem[Graham and Steen, 1994]{Graham1994}
Graham, M.~D. and Steen, P.~H. (1994).
\newblock Plume formation and resonant bifurcations in porous-media convection.
\newblock {\em Journal of Fluid Mechanics}, 272(-1):67.

\bibitem[Hamadouche et~al., 2016]{Hamadouche2016}
Hamadouche, A., Nebbali, R., Benahmed, H., Kouidri, A., and Bousri, A. (2016).
\newblock Experimental investigation of convective heat transfer in an
  open-cell aluminum foams.
\newblock {\em Experimental Thermal and Fluid Science}, 71:86--94.

\bibitem[Hewitt et~al., 2012]{Hewitt2012}
Hewitt, D.~R., Neufeld, J.~A., and Lister, J.~R. (2012).
\newblock Ultimate regime of high {R}ayleigh number convection in a porous
  medium.
\newblock {\em Physical Review Letters}, 108(22).

\bibitem[Hewitt et~al., 2013a]{Hewitt2013a}
Hewitt, D.~R., Neufeld, J.~A., and Lister, J.~R. (2013a).
\newblock Convective shutdown in a porous medium at high {Rayleigh} number.
\newblock {\em J. Fluid Mech.}, 719:551--586.

\bibitem[Hewitt et~al., 2013b]{Hewitt2013b}
Hewitt, D.~R., Neufeld, J.~A., and Lister, J.~R. (2013b).
\newblock Stability of columnar convection in a porous medium.
\newblock {\em Journal of Fluid Mechanics}, 737:205–231.

\bibitem[Hidalgo and Carrera, 2009]{Hidalgo2009}
Hidalgo, J.~J. and Carrera, J. (2009).
\newblock Effect of dispersion on the onset of convection during co2
  sequestration.
\newblock {\em Journal of Fluid Mechanics}, 640:441--452.

\bibitem[Hidalgo et~al., 2015]{Hidalgo2015}
Hidalgo, J.~J., Dentz, M., Cabeza, Y., and Carrera, J. (2015).
\newblock Dissolution patterns and mixing dynamics in unstable reactive flow.
\newblock {\em Geophysical Research Letters}, 42(15):6357–6364.

\bibitem[Hidalgo et~al., 2012]{Hidalgo2012}
Hidalgo, J.~J., Fe, J., Cueto-Felgueroso, L., and Juanes, R. (2012).
\newblock Scaling of convective mixing in porous media.
\newblock {\em Phys. Rev. Lett.}, 109(26).

\bibitem[Hidalgo et~al., 2013]{Hidalgo2013}
Hidalgo, J.~J., MacMinn, C.~W., and Juanes, R. (2013).
\newblock Dynamics of convective dissolution from a migrating current of carbon
  dioxide.
\newblock {\em Ad. Water Resour.}, 62:511--519.

\bibitem[Horton and Rogers, 1945]{Horton1945}
Horton, C.~W. and Rogers, F.~T. (1945).
\newblock Convection currents in a porous medium.
\newblock {\em Journal of Applied Physics}, 16(6):367–370.

\bibitem[Howard, 1966]{Howard1966}
Howard, L.~N. (1966).
\newblock {\em Convection at high Rayleigh number}, pages 1109--1115.
\newblock Springer Berlin Heidelberg, Berlin, Heidelberg.

\bibitem[Kimura et~al., 1986]{Kimura1986}
Kimura, S., Schubert, G., and Straus, J.~M. (1986).
\newblock Route to chaos in porous-medium thermal convection.
\newblock {\em Journal of Fluid Mechanics}, 166(-1):305 -- 324.

\bibitem[Kitanidis, 1994]{Kitanidis:1994}
Kitanidis, P.~K. (1994).
\newblock The concept of the dilution index.
\newblock {\em Water Resour. Res.}, 30(7):2011--2026.

\bibitem[Kueper and Frind, 1991]{Kueper1991}
Kueper, B.~H. and Frind, E.~O. (1991).
\newblock Two-phase flow in heterogeneous porous media: 1. model development.
\newblock {\em Water Resources Research}, 27(6):1049--1057.

\bibitem[Lapwood, 1948]{Lapwood1948}
Lapwood, E.~R. (1948).
\newblock Convection of a fluid in a porous medium.
\newblock {\em Mathematical Proceedings of the Cambridge Philosophical
  Society}, 44(04):508.

\bibitem[Le~Borgne et~al., 2010]{LeBorgne2010}
Le~Borgne, T., Dentz, M., Bolster, D., Carrera, J., de~Dreuzy, J.-R., and Davy,
  P. (2010).
\newblock Non-fickian mixing: Temporal evolution of the scalar dissipation rate
  in heterogeneous porous media.
\newblock {\em Ad. Water Resour.}, 33(12):1468--1475.

\bibitem[Le~Borgne et~al., 2013]{LeBorgne2013}
Le~Borgne, T., Dentz, M., and Villermaux, E. (2013).
\newblock Stretching, coalescence, and mixing in porous media.
\newblock {\em Phys, Rev. Lett.}, 110(20).

\bibitem[Le~Borgne et~al., 2015]{LeBorgneDentzVillermaux2015}
Le~Borgne, T., Dentz, M., and Villermaux, E. (2015).
\newblock The lamellar description of mixing in porous media.
\newblock {\em J. Fluid Mech,}, 770:458--498.

\bibitem[Martin et~al., 1987]{Martin1987}
Martin, D., Griffiths, R.~W., and Campbell, I.~H. (1987).
\newblock Compositional and thermal convection in magma chambers.
\newblock {\em Contributions to Mineralogy and Petrology}, 96(4):465--475.

\bibitem[Musgrave, 1985]{Musgrave1985}
Musgrave, D.~L. (1985).
\newblock A numerical study of the roles of subgyre-scale mixing and the
  western boundary current on homogenization of a passive tracer.
\newblock {\em Journal of Geophysical Research: Oceans}, 90(C4):7037--7043.

\bibitem[Neufeld et~al., 2010]{Neufeld2010}
Neufeld, J.~A., Hesse, M.~A., Riaz, A., Hallworth, M.~A., Tchelepi, H.~A., and
  Huppert, H.~E. (2010).
\newblock Convective dissolution of carbon dioxide in saline aquifers.
\newblock {\em Geophys. Res. Lett.}, 37(22):L22404.

\bibitem[Otero et~al., 2004]{Otero2004}
Otero, J., Dontcheva, L.~A., Johnston, H., Worthing, R.~A., Kurganov, A.,
  Petrova, G., and Doering, C.~R. (2004).
\newblock High-{Rayleigh}-number convection in a fluid-saturated porous layer.
\newblock {\em Journal of Fluid Mechanics}, 500:263–281.

\bibitem[Ottino, 1989]{Ottino1989}
Ottino, J. (1989).
\newblock {\em The Kinematics of Mixing: Stretching, Chaos, and Transport}.
\newblock Cambridge Texts in Applied Mathematics. Cambridge University Press.

\bibitem[Ranz, 1979]{Ranz1979}
Ranz, W.~E. (1979).
\newblock Applications of a stretch model to mixing, diffusion, and reaction in
  laminar and turbulent flows.
\newblock {\em AIChE J.}, 25(1):41--47.

\bibitem[Rees et~al., 2008]{Rees2008}
Rees, D., Selim, A., and Ennis-King, J. (2008).
\newblock The instability of unsteady boundary layers in porous media.
\newblock In Vadasz, P., editor, {\em Emerging Topics in Heat and Mass Transfer
  in Porous Media. From Bioengineering and Microelectronics to Nanotechnology},
  volume~22 of {\em Theory and Applications of Transport in Porous Media},
  pages 85--110. Springer, Netherlands.

\bibitem[Riaz et~al., 2006]{Riaz2006}
Riaz, A., Hesse, M., Tchelepi, H.~A., and {Orr Jr.}, F.~M. (2006).
\newblock Onset of convection in a gravitationally unstable diffusive boundary
  layer in porous media.
\newblock {\em J. Fluid Mech.}, 548:87--111.

\bibitem[Sanford et~al., 1998]{Sanford1998}
Sanford, W.~E., Whitaker, F.~F., Smart, P.~L., and Jones, G. (1998).
\newblock Numerical analysis of seawater circulation in carbonate platforms: I,
  geothermal convection.
\newblock {\em American Journal of Science}, 298(10):801--828.

\bibitem[Shadden et~al., 2005]{Shadden2005}
Shadden, S.~C., Lekien, F., and Marsden, J.~E. (2005).
\newblock Definition and properties of lagrangian coherent structures from
  finite-time {Lyapunov} exponents in two-dimensional aperiodic flows.
\newblock {\em Physica D: Nonlinear Phenomena}, 212(3-4):271–304.

\bibitem[Slim, 2014]{Slim2014}
Slim, A.~C. (2014).
\newblock Solutal-convection regimes in a two-dimensional porous medium.
\newblock {\em Journal of Fluid Mechanics}, 741:461–491.

\bibitem[Slim and Ramakrishnan, 2010]{Slim2010}
Slim, A.~C. and Ramakrishnan, T.~S. (2010).
\newblock Onset and cessation of time-dependent, dissolution-driven convection
  in porous media.
\newblock {\em Physics of Fluids}, 22(12):124103.

\bibitem[Szulczewski et~al., 2013]{Szulczewski2013}
Szulczewski, M.~L., Hesse, M.~A., and Juanes, R. (2013).
\newblock Carbon dioxide dissolution in structural and stratigraphic traps.
\newblock {\em Journal of Fluid Mechanics}, 736:287--315.

\bibitem[Tait and Jaupart, 1989]{Tait1989}
Tait, S. and Jaupart, C. (1989).
\newblock Compositional convection in viscous melts.
\newblock {\em Nature}, 338(6216):571--574.

\bibitem[Villermaux, 2012]{Villermaux2012}
Villermaux, E. (2012).
\newblock Mixing by porous media.
\newblock {\em C. R. Mecanique}, 340(11-12):933--943.

\bibitem[Villermaux and Duplat, 2006]{Villermaux2006}
Villermaux, E. and Duplat, J. (2006).
\newblock Coarse grained scale of turbulent mixtures.
\newblock {\em Phys. Rev. Lett.}, 97:144506.

\bibitem[Wells et~al., 2011]{Wells2011}
Wells, A.~J., Wettlaufer, J.~S., and Orszag, S.~A. (2011).
\newblock Brine fluxes from growing sea ice.
\newblock {\em Geophysical Research Letters}, 38(4).
\newblock L04501.

\end{thebibliography}
\bibliographystyle{apalike}
\end{document}